\documentclass[10pt]{article}

\usepackage[a4paper,top=2.5cm,bottom=2.5cm,left=2.5cm,right=2.5cm,marginparwidth=1.75cm]{geometry}
\usepackage{amssymb}
\usepackage{amsmath}
\usepackage{amsthm}
\usepackage{amsfonts}
\usepackage{mathtools}
\usepackage{graphicx}
\usepackage{float}
\usepackage{color}
\usepackage{fullpage}
\usepackage[normalem]{ulem} 
\usepackage{makeidx}
\usepackage{xspace}
\usepackage{authblk}
\usepackage{datetime}
\usepackage{microtype}
\usepackage{xpatch}
\usepackage{xstring}
\usepackage{fancyhdr}
\usepackage{mathrsfs}
\usepackage{esint}
\usepackage{upgreek}
\usepackage{siunitx}
\usepackage{tabularx, multirow}
\usepackage{threeparttable}
\usepackage[]{booktabs} 
\usepackage{float}
\usepackage[colorlinks=false]{hyperref}
\usepackage[noabbrev]{cleveref}
\usepackage[font = footnotesize, labelfont=bf]{caption}

\fancypagestyle{fpg}{%
	\lfoot{\footnotesize Journal of Physics D: Applied Physics \textbf{2020}}%
	\cfoot{\normalsize\thepage}%
	\rfoot{\footnotesize \href{https://doi.org/10.1088/1361-6463/abca62}{DOI: 10.1088/1361-6463/abca62}}%
}

\makeindex

\begin{document}

\title{\Large\textbf{Numerical Study of Molten Metal Melt Pool Behaviour during Conduction-mode Laser Spot Melting}}
 
	\author[1,$\dagger$]{Amin Ebrahimi}
	\author[2]{Chris R. Kleijn}
	\author{Ian M. Richardson}
	\affil[1]{\small \textit{Department of Materials Science and Engineering, Delft University of Technology, Mekelweg~2, 2628CD~Delft, The~Netherlands}}
	\affil[2]{\small \textit{Department of Chemical Engineering, Delft University of Technology, van~der~Maasweg~9, 2629HZ Delft, The~Netherlands \vspace{5mm}}}
	\affil[$\dagger$]{\small Corresponding author, Email: A.Ebrahimi@tudelft.nl; Ebrahimi.Amin@gmail.com}
	
\date{}
\maketitle
\thispagestyle{fpg}

\begin{abstract}
		Molten metal melt pools are characterised by highly non-linear responses, which are very sensitive to imposed boundary conditions. Temporal and spatial variations in the~energy flux distribution are often neglected in numerical simulations of melt pool behaviour. Additionally, thermo-physical properties of materials are commonly changed to achieve agreement between predicted melt-pool shape and experimental post-solidification macrograph. Focusing on laser spot melting in conduction~mode, we investigated the~influence of dynamically adjusted energy flux distribution and changing thermo-physical material properties on melt pool oscillatory behaviour using both deformable and non-deformable assumptions for the~gas-metal interface. Our results demonstrate that adjusting the~absorbed energy flux affects the~oscillatory fluid flow behaviour in the~melt pool and consequently the~predicted melt-pool shape and size. We~also~show that changing the~thermo-physical material properties artificially or using a~non-deformable surface assumption lead to significant differences in melt pool oscillatory behaviour compared to the~cases in which these assumptions are not made.
\end{abstract}

\newpage
\section{Introduction}
\label{sec:intro}

Laser melting is being utilised for material processing such as additive manufacturing, joining, cutting and surface modification. The~results of experimentation performed by~Ayoola \textit{et~al.}~\cite{Ayoola_2017} revealed that the~energy flux distribution over the~melt-pool surface can affect melting, convection and energy transport in liquid melt pools and the~subsequent re-solidification during laser melting processes. The~imposed energy flux heats and melts the~material and generates temperature gradients over the~melt-pool surface. The~resulting surface tension gradients and therefore Marangoni force is often the~dominant force driving fluid flow, as can be understood virtually from the~numerical investigation conducted by~\mbox{Oreper and Szekely~\cite{Oreper_1984}} and experimental observations reported by~Mills \textit{et~al.}~\cite{Mills_1998}. Experimental investigations of~\mbox{Heiple and Roper~\cite{Heiple_1982}} showed that the~presence of surfactants in molten materials can alter Marangoni convection in the~melt pool, and thus the~melt-pool shape. Moreover, Paul and DebRoy~\cite{Paul_1988} reported that the~smoothness of the~melt pool surface decreases when surfactants are present in the~melt pool. However, according to the~literature survey conducted by~Cook and Murphy~\cite{Cook_2020}, the~influence of surfactants on variations of surface tension and its temperature gradient is often neglected in numerical simulations of welding and additive manufacturing. \mbox{DebRoy and David~\cite{DebRoy_1995}} stated that fluid flow in melt pools can lead to deformation and oscillation of the~liquid free surface and can affect the~stability of the~process and the~structure and properties of the~solid materials after re-solidification. \mbox{Aucott \textit{et~al.}~\cite{Aucott_2018}} confirmed that transport phenomena during fusion welding and additive manufacturing processes are characterised by highly non-linear responses that are very sensitive to material composition and imposed heat flux boundary~conditions. Numerical models capable of predicting the~melt pool behaviour with a~sufficient level of accuracy are thus required to gain an~insight into the~physics of fluid flow and the~nature of flow instabilities that are not accessible experimentally.

To avoid excessive simulation complexity and execution time in numerical simulations, the~melt pool in such simulations is often decoupled from the~heat source, the~latter being incorporated as a~boundary condition at the~melt-pool surface. These boundary conditions should be imposed with a~sufficient level of accuracy, as it is known from the~work of \mbox{Zacharia \textit{et~al.}~\cite{Zacharia_1990}} that modelling the~interfacial phenomena is critical to predicting the~melt pool behaviour. Numerical studies on melt pool behaviour reported in the~literature use both deformable and non-deformable surface assumptions for the~gas-metal interface. Comparing the~melt-pool shapes predicted using both deformable and non-deformable surface assumptions, \mbox{Ha and Kim~\cite{Ha_2005}} concluded that free-surface oscillations can enhance convection in the~melt pool and influence the~melt-pool shape. Shah \textit{et~al.}~\cite{Shah_2018} reported that the~difference between melt-pool shapes obtained from numerical simulations with deformable and non-deformable surface assumptions depends on the~laser power. Three-dimensionality of the~molten metal flow in melt pools, as~observed experimentally by \mbox{Zhao \textit{et~al.}~\cite{Zhao_2010}} and numerically by Kidess \textit{et~al.}~\cite{Kidess_2016}, is often neglected in numerical simulations. Moreover, when accounting for surface deformations, the~volume-of-fluid (VOF) method developed by \mbox{Hirt and Nichols~\cite{Hirt_1981}}, based on a~Eulerian formulation, is the~most common method for modelling the~melt~pool behaviour. In this diffuse boundary method, the~interfacial forces and the~energy fluxes applied on the~melt-pool surface are treated as volumetric source terms in the~surface region, instead of imposing them as boundary conditions. In~this approach, however, the~fact that surface deformations lead to temporal and spatial variations of the~free~surface boundary conditions, as remarked by Meng \textit{et~al.}~\cite{Meng_2016} and \mbox{Wu \textit{et~al.}~\cite{Wu_2019}}, is often neglected. The~results reported by \mbox{Choo \textit{et~al.}~\cite{Choo_1990}} suggest that variations in power-density distribution and changes in free-surface profile can affect molten metal flow in melt pools and its stability. Further investigations are essential to improve the~understanding of the~complex transport phenomena that happen during laser spot melting.

The aim of the~present work is to analyse the~effects of a~dynamically adjusting energy flux distribution over the~deforming liquid surface on the~nature of fluid flow instabilities in partially-penetrated liquid melt pools. We will particularly focus on flow instabilities in low-Prandtl number liquid metal pools during conduction-mode laser spot melting; our results should however be relevant for a~much wider range of materials processing technologies. Three-dimensional calculations are carried out to numerically predict the~melt pool behaviour and thermocapillary-driven flow instabilities using various heat source implementation methods. Our study provides a~quantitative representation and an~understanding of the~influence of heat flux boundary conditions on the~transport phenomena and flow instabilities in the~molten metal melt pool. Additionally, we discuss the~influence of artificially enhanced transport coefficients on the~melt-pool oscillatory behaviour.

\section{Model description}
\label{sec:model}

\subsection{Physical model}
\label{sec:phys_model}

\begin{figure}[H] 
	\centering
	\includegraphics[width=0.75\linewidth]{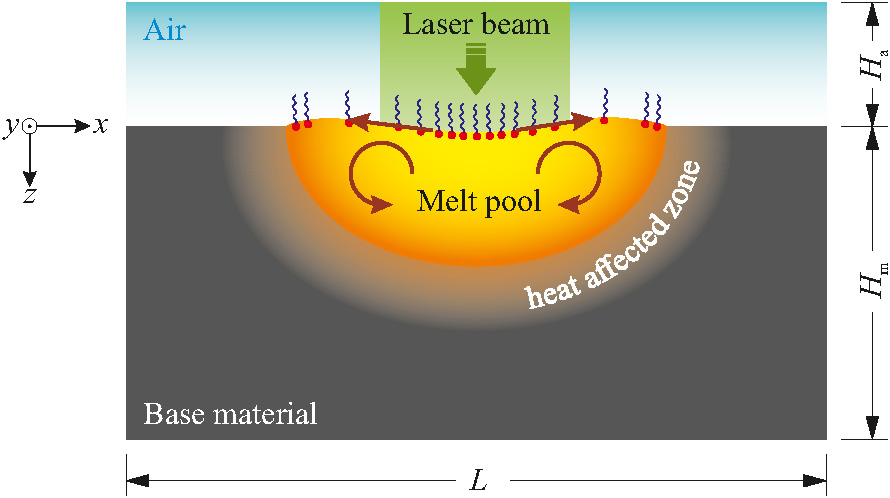}
	\caption{Schematic of a~conduction-mode laser spot melting.}
	\label{fig:schematic}
\end{figure}

Laser spot melting of a~metallic S705 alloy, as shown schematically in \cref{fig:schematic} and as experimentally studied by \mbox{Pitscheneder \textit{et~al.}~\cite{Pitscheneder_1996}}, was numerically simulated as a~representative example in the~present work. A~defocused laser-beam with a~radius of $r_\mathrm{b} = \SI{1.4}{\milli\meter}$ heats the~bulk material from its top surface for~$\SI{4}{\second}$. The~laser-beam power $Q$ is set to $\SI{3850}{\watt}$ with a~top-hat intensity distribution. The~averaged laser absorptivity of the~material surface $\eta$ is assumed to remain constant at $13\%$~\cite{Pitscheneder_1996}. The~absorbed laser power leads to an~increase in temperature and subsequent melting of the~base material. The~base material is a~rectangular cuboid shape with a~base size ($L\times L$) of $24 \times \SI{24}{\milli\meter\squared}$ and a~height ($H_\mathrm{m}$) of $\SI{10}{\milli\meter}$, initially at an~ambient temperature of $T_\mathrm{i} = \SI{300}{\kelvin}$. a~layer of air with a~thickness of $H_\mathrm{a} = \SI{2}{\milli\meter}$ is considered above the~base material to monitor the~gas-metal interface evolution. Except for the~surface tension, the~material properties are assumed to be constant and temperature independent and are presented in \cref{tab:material_properties}. Although~temperature-dependent properties can be employed in the~present model to enhance the~model accuracy, employing temperature-independent properties facilitates comparison of the~results of the~present work with previous results published in the~literature \cite{Saldi_2013,Kidess_2016_PhysFlu,Ha_2005}, where temperature-independent material properties are employed. Moreover, further studies are required to enhance the~accuracy of calculation and measurement of temperature-dependent material properties, particularly for the~liquid phase above the~melting temperature. The~effects of employing temperature-dependent material properties on the~thermal and fluid flow fields and melt-pool shape are discussed in detail in \cref{sec:the_effects_of_temperature_dependent_properties}. A~change in surface-tension due to the~non-uniform temperature distribution over the~gas-liquid interface induces thermocapillary stresses that drive the~melt~flow. This fluid motion from low to high surface-tension regions changes the~temperature distribution in the~melt pool \cite{Heiple_1983} and can lead to surface deformations that change the~melt-pool shape and properties of the~material after solidification \cite{Mills_1990}. The~sulphur contained in the~alloy can alter the~surface tension of the~molten material and its variations with temperature \mbox{\cite{Sahoo_1988}}. 

\begin{table}[H] 
	\centering
	\caption{Thermophysical properties of the~Fe-S alloy and air used in the~present study. Values are taken from \cite{Kidess_2016_PhysFlu}.}
	\begin{tabular}{llll}
		\hline
		Property                              & Fe-S alloy  & Air          & Unit                                \\ \hline
		Density $\rho$                        & \SI{8100}   & \SI{1.225}   & \si{\kilogram\per\meter\cubed}      \\
		Specific heat capacity $c_\mathrm{p}$ & \SI{670}    & \SI{1006}    & \si{\joule\per\kilogram\per\kelvin} \\
		Thermal conductivity $k$              & \SI{22.9}   & \SI{0.024}   & \si{\watt\per\meter\kelvin}         \\
		Viscosity $\mu$                       & \SI{6e-3}   & \SI{1.8e-05} & \si{\kilogram\per\meter\per\second} \\
		Latent heat of fusion $L_\mathrm{f}$  & \SI{250800} & --           & \si{\joule\per\kilogram}            \\
		Liquidus temperature $T_\mathrm{l}$   & \SI{1620}   & --           & \si{\kelvin}                        \\
		Solidus temperature $T_\mathrm{s}$    & \SI{1610}   & --           & \si{\kelvin}                        \\ \hline
	\end{tabular} 
	\label{tab:material_properties}
\end{table}

\subsection{Mathematical models}
\label{sec:math_model}

A three-dimensional multiphase model based on the~finite-volume method was utilised to predict the~melt pool dynamic behaviour and the~associated transport phenomena. {The present model is developed for conduction-mode laser melting. Further considerations will be required to develop the~model for keyhole mode welding, including the~complex laser-matter interactions, changes in surface chemistry and consequent surface tension variations for non- or partially shielded welding conditions.} Both the~molten metal and air were treated as Newtonian and incompressible fluids. The~rmal buoyancy forces, which in laser spot melting are generally negligible compared to thermocapillary forces \cite{Zacharia_1989}, were neglected both in the~liquid metal and in the~air.
During conduction-mode laser-melting, the~heat-source energy density is too low to cause significant vaporisation of the~liquid metal \cite{Cho_2015}, and the~surface deformations are too small to cause multiple reflections of the~laser rays at the~liquid surface \cite{Kim_2008}. {Furthermore, our preliminary results indicate that the~maximum temperature of the~melt pool is predominantly below $\SI{2700}{\kelvin}$, significantly less than the~boiling temperature of stainless steels, which is typically $\mathcal{O}(3100)\, \SI{}{\kelvin}$.} Hence, multiple reflections and vaporisation were ignored. With this, the~unsteady conservation equations of mass, momentum and energy were defined as follows:

\begin{equation}
	\nabla\cdot\left(\rho\vec{V}\right) = 0,
	\label{eq: mass}
\end{equation}

\begin{equation}
	\frac{\partial}{\partial t} \left(\rho \vec{V}\right) + \nabla\cdot\left(\rho\vec{V} \vec{V}\right) = 
	-\nabla p + \nabla\cdot\left(\mu\nabla\vec{V}\right) + \vec{S}_\mathrm{d} + \vec{F}_\mathrm{s},
	\label{eq: momentum}
\end{equation}

\begin{equation}
	\frac{\partial}{\partial t} \left(\rho h\right) + \nabla\cdot\left(\rho\vec{V} h\right) = 
	\nabla\cdot\left(\frac{k}{c_\mathrm{p}}\nabla h\right) - \frac{\partial}{\partial t} \left(\rho \Delta H\right) - \nabla\cdot\left(\rho\vec{V}\Delta H\right) + S_\mathrm{T},
	\label{eq: energy}
\end{equation}

\noindent
where, $\rho$ is density, $t$ time, $\vec{V}$ velocity vector, $p$ pressure, $\mu$ dynamic viscosity, $h$ sensible heat, $k$ thermal conductivity, $c_\mathrm{p}$ specific heat capacity at constant pressure and $\Delta H$ latent heat. The~enthalpy of the~material $H$ was defined as the~sum of the~latent heat and the~sensible heat \cite{Voller_1991}, and is expressed as

\begin{equation}
	\begingroup
	H = h + \Delta H = \left(h_\mathrm{ref} + \int_{T_\mathrm{ref}}^{T} c_\mathrm{p} \mathop{}\!\mathrm{d}T\right) + f_\mathrm{L}L_\mathrm{f},
	\endgroup
	\label{eq: enthalpy}
\end{equation}

\noindent
where, $h_\mathrm{ref}$ is reference enthalpy, $T_\mathrm{ref}$ reference temperature, $L_\mathrm{f}$ latent heat of fusion, and $f_\mathrm{L}$ local liquid volume-fraction. Assuming the~liquid volume-fraction in the~metal to be a~function of temperature only, a~linear function \cite{Voller_1991} was used to calculate the~liquid volume-fraction as follows:

\begin{equation}
	f_\mathrm{L} = \frac{T - T_\mathrm{s}}{T_\mathrm{l} - T_\mathrm{s}}; \quad T_\mathrm{s} \le T \le T_\mathrm{l},
	\label{eq: liquid_fraction}
\end{equation}

\noindent
where, $T_\mathrm{s}$ and $T_\mathrm{l}$ are the~solidus and liquidus temperatures, respectively.

To suppress fluid velocities in solid regions and to model fluid flow damping in the~so-called  mushy zone, where phase transformation occurs within a~temperature range between $T_\mathrm{s}$ and $T_\mathrm{l}$, a~momentum sink term $\vec{S}_\mathrm{d}$ \cite{Voller_1987} was implemented into the~momentum equation as

\begin{equation}
	\vec{S}_\mathrm{d} = C\ \frac{(1 - f_\mathrm{L})^2}{f_\mathrm{L}^3 + \epsilon} \ \vec{V},
	\label{eq: momentum_sink}
\end{equation}

\noindent
where, $C$ is the~mushy-zone constant and $\epsilon$ is a~small number to avoid division by zero as $f_\mathrm{L}$ approaches 0. The~mushy-zone constant $C$ was set to $10^7\,\SI{}{\kilogram\per\square\meter\per\square\second}$, which is large enough to dampen fluid velocities in solid regions according to the~criterion defined by Ebrahimi \textit{et~al.}~\cite{Ebrahimi_2019}, and $\epsilon$ was equal to $10^{-3}$. 

To capture the~position of the~gas-metal interface during melting, the~volume-of-fluid (VOF) method developed by~Hirt and Nichols~\cite{Hirt_1981} was utilised. This requires the~solution of a~transport equation for one additional scalar variable $\phi$ (\textit{i.e.},~the so-called volume-of-fluid fraction) that varies between 0 in the~gas phase and 1 in the~metal phase:

\begin{equation}
	\frac{\partial \phi}{\partial t} + \nabla\cdot\left(\phi\vec{V}\right) = 0.
	\label{eq: vof}
\end{equation}

\noindent
Computational cells with $0 \le \phi \le 1$ are in the~gas-metal interface region. The~effective thermophysical properties of the~fluid is then calculated using the~mixture model as follows \cite{Sun_2004}:

\begin{equation}
	\begingroup
	\zeta = \phi \, \zeta_\mathrm{metal} + \left(1-\phi\right) \zeta_\mathrm{gas},
	\endgroup
	\label{eq: mixture_model}
\end{equation}

\noindent
where, $\zeta$ corresponds to density $\rho$, viscosity $\mu$, thermal conductivity $k$ and specific heat capacity $c_\mathrm{p}$.

Based on the~method developed by~Brackbill \textit{et~al.}~\cite{Brackbill_1992}, surface tension and thermocapillary forces acting on the~gas-metal interface were applied as volumetric forces in these interface cells and were introduced into the~momentum equation as a~source term $\vec{F}_\mathrm{s}$ as follows:

\begin{equation}
	\vec{F}_\mathrm{s} = \vec{f}_\mathrm{s} \lVert \nabla\phi \rVert\frac{2\rho}{\rho_\mathrm{gas} + \rho_\mathrm{metal}},
	\label{eq: surface_force_source}
\end{equation}

\noindent
where subscripts indicate the~phase. The~term $2\rho/\!\left(\rho_\mathrm{gas} + \rho_\mathrm{metal}\right)$ was utilised to abate the~effect of the~large metal-to-gas density ratio by redistributing the~volumetric surface-forces towards the~metal phase (\textit{i.e.}~the~heavier phase). $\vec{f}_\mathrm{s}$ is the~surface force per unit area and was defined as follows:

\begin{equation}
	\vec{f}_\mathrm{s} = \sigma\kappa\vec{n} + \frac{\mathrm{d}\sigma}{\mathrm{d}T} \left[\nabla T - \vec{n}\left(\vec{n}\cdot\nabla T\right)\right],
	\label{eq: surface_force}
\end{equation}

\noindent
where, {the first term on the~right-hand side is the~normal component, which depends on the~value of surface tension and curvature of the~melt-pool surface, and the~second term is the~tangential component that is affected by the~surface tension gradient. In \cref{eq: surface_force},} $\sigma$ is surface tension, and $\kappa$ the~surface curvature is

\begin{equation}
	\kappa = -\left(\nabla\cdot\vec{n}\right),
	\label{eq: curvature}
\end{equation}

\noindent
and $\vec{n}$ the~surface unit normal vector is

\begin{equation}
	\vec{n} = \frac{\nabla\phi}{\lVert \nabla \phi \rVert}.
	\label{eq: normal}
\end{equation}

\subsubsection{Surface tension}
\label{sec:surface_tension}

\begin{figure}[H] 
	\centering
	\includegraphics[width=0.55\linewidth]{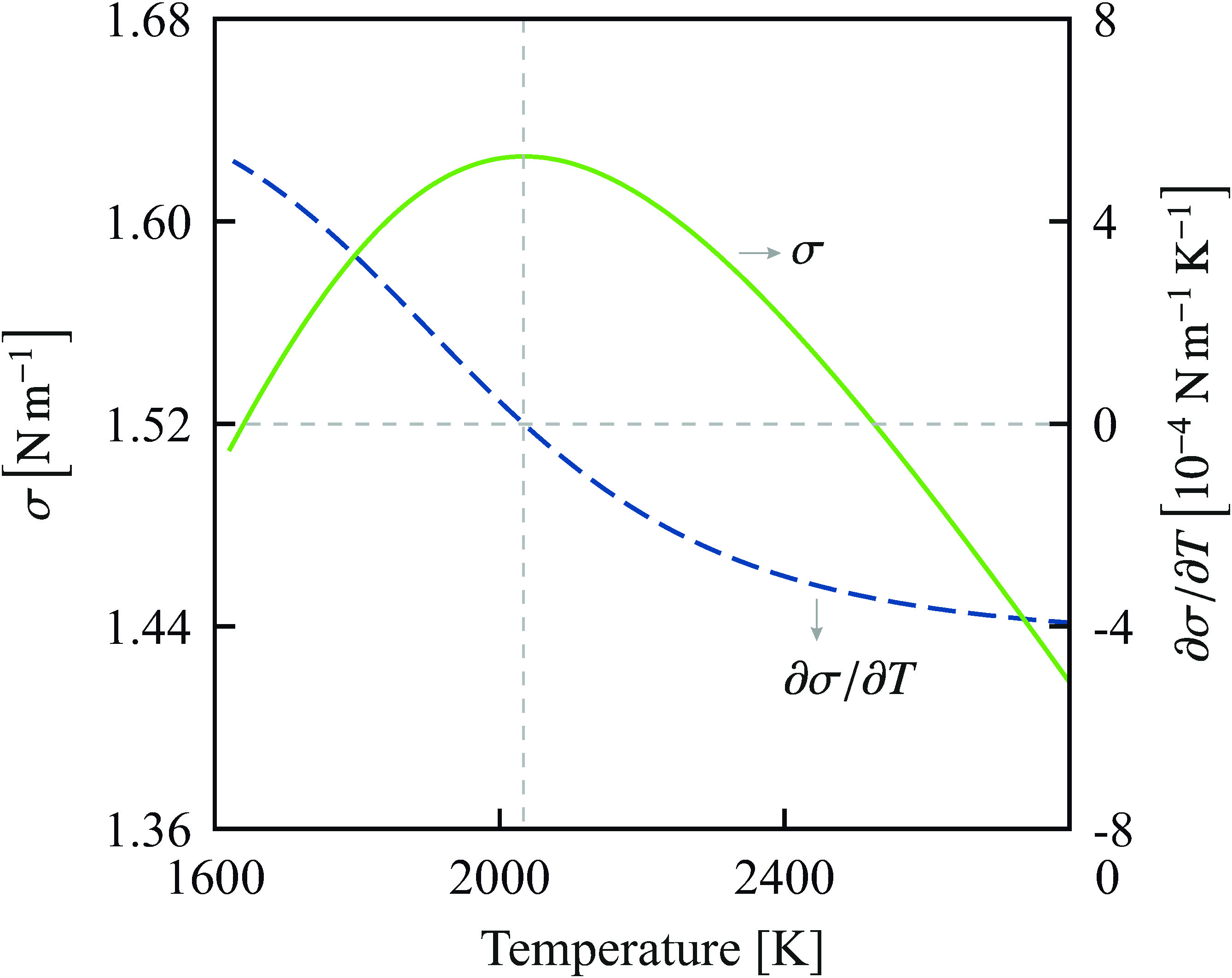}
	\caption{Variations of surface tension (green solid-line) and its temperature gradient (blue dashed-line) as a~function of temperature obtained from \cref{eq: surface_tension_FeS} for an~Fe-S alloy containing $\SI{150}{ppm}$ sulphur.}
	\label{fig:surface_tension_variations}
\end{figure}

The temperature dependence of the~surface tension of a~liquid solution with a~low concentration of surfactant can be approximated using a~theoretical correlation derived on the~basis of the~combination of Gibbs and Langmuir adsorption isotherms as follows \cite{Belton_1976,March_1991}:

\begin{equation}
	\sigma = \sigma^\circ - \mathrm{R}\, T\, \Gamma_\mathrm{s}\, \ln(1 + K a_\mathrm{s}),
	\label{eq: surface_tension}
\end{equation}

\noindent
where, $\sigma$ is the~surface tension of the~solution, $\sigma^\circ$ the~pure solvent surface tension, R the~gas constant, $\Gamma_\mathrm{s}$ the~adsorption at saturation, $K$ the~adsorption coefficient, and $a_\mathrm{s}$ the~activity of the~solute. From this, \mbox{Sahoo \textit{et~al.}~\cite{Sahoo_1988}} derived a~correlation for binary molten metal-surfactant systems, including binary Fe-S systems:

\begin{equation}
	\sigma = \sigma_\mathrm{m}^\circ + \left(\frac{\partial\sigma}{\partial T}\right)^\circ \left(T - T_\mathrm{m}\right) - \mathrm{R}\, T\, \Gamma_\mathrm{s}\, \ln\!\left[1 + \psi\, a_\mathrm{s} \exp\!\left(\frac{-\Delta H^\circ}{\mathrm{R} T}\right)\right],
	\label{eq: surface_tension_FeS}
\end{equation}

\noindent
where $\sigma_\mathrm{m}^\circ$ is the~surface tension of pure molten-metal at the~melting temperature $T_\mathrm{m}$, $\psi$ an~entropy factor, $\Delta H^\circ$ the~standard heat of adsorption, and $\left(\partial \sigma / \partial T\right)^\circ$ the~temperature coefficient of the~surface tension of the~pure molten-metal. Values of the~properties used in \cref{eq: surface_tension_FeS} to calculate the~temperature dependence of the~surface tension of the~molten Fe-S alloy are presented in \cref{tab:surface_tension_FeS}. Variations of the~surface tension and its temperature coefficient are shown in \cref{fig:surface_tension_variations} for an~Fe-S alloy with $\SI{150}{ppm}$ sulphur content.

\begin{table}[H] 
	\centering
	\caption{The values used to calculate the~surface tension of molten Fe-S alloy \cite{Sahoo_1988}.}
	\begin{tabular}{lll}
		\hline
		Property                                          & Value         & Unit                              \\ \hline
		$\sigma_\mathrm{m}^\circ$                         & \SI{1.943}    & \si{\newton\per\meter}            \\
		$\left(\partial \sigma / \partial T\right)^\circ$ & \SI{-4.3e-4}  & \si{\newton\per\meter\per\kelvin} \\
		$\Gamma_\mathrm{s}$                               & \SI{1.3e-5}   & \si{\mole\per\square\meter}       \\
		$\psi$                                            & \SI{3.18e-3}  & --                                \\
		$a_\mathrm{s}$                                    & \SI{150}      & --                                \\
		$\Delta H^\circ$                                  & \SI{-166.2e5} & \si{\joule\per\mole}              \\ \hline
	\end{tabular} 
	\label{tab:surface_tension_FeS}
\end{table}

\subsubsection{Laser heat source}
\label{sec:laser_heat_source}

The laser supplies a~certain amount of energy ($\eta\, Q$) to the~material. Its energy input was modelled by adding a~volumetric source term $S_\mathrm{T}$ to the~energy equation (\cref{eq: energy}) in cells at the~gas-metal interface. The~surface deformations affect the~total energy input to the~material \cite{Gao_2017}. This effect is generally not considered in numerical simulations of melt pool behaviour. To investigate the~influence of surface deformations, five different methods for the~heat-source implementation were considered.

\paragraph{Case~1, heat-source implementation without taking the~influence of surface deformations on total energy input into consideration}
In this case, the~heat source model is defined as

\begin{equation}
	S_\mathrm{T} =
	\begin{dcases}
		\frac{\eta\, Q}{\pi r_\mathrm{b}^2} \lVert \nabla\phi \rVert\frac{2\,\rho\, c_\mathrm{p}}{\left(\rho\, c_\mathrm{p}\right)_\mathrm{gas} + \left(\rho\, c_\mathrm{p}\right)_\mathrm{metal}}, & \text{if $r \le r_\mathrm{b}$} \\
		0, & \text{otherwise}
	\end{dcases}
	\label{eq: heat_source_case1}
\end{equation}

\noindent
where, $r$ is radius defined as $\sqrt{x^2 + y^2}$.	This method is the~most common method in modelling melt-pool surface oscillations without having a~deep penetration (keyhole formation) (see for instance, \cite{Cho_2015,Kim_2008}). However, variations in the~total energy input caused by a~change in melt-pool surface shape, through changes in $\lVert \nabla\phi \rVert$, causing $\iiint \limits_{\mathrm{\forall}} S_\mathrm{T} \mathop{}\!\mathrm{d}V$ to be different from $\eta Q$, are an~inherent consequence of using this method. Additionally, every segment of the~melt-pool surface is exposed to the~same amount of energy input regardless of the~local surface orientation.

\paragraph{Case~2, heat source adjustment}
To conserve the~total energy input, an~adjustment coefficient $\xi$ was introduced to the~heat source model as follows:

\begin{equation}
	S_\mathrm{T} =
	\begin{dcases}
		\frac{\xi\, \eta\, Q}{\pi r_\mathrm{b}^2} \lVert \nabla\phi \rVert\frac{2\,\rho\, c_\mathrm{p}}{\left(\rho\, c_\mathrm{p}\right)_\mathrm{gas} + \left(\rho\, c_\mathrm{p}\right)_\mathrm{metal}}, & \text{if $r \le r_\mathrm{b}$} \\
		0, & \text{otherwise}
	\end{dcases}
	\label{eq: heat_source_case2}
\end{equation}

\noindent where, $\xi$ was evaluated at the~beginning of every time-step and was defined as
\begin{equation}
	\xi = \frac{\eta \, Q}{\iiint \limits_{\mathrm{\forall}} S_\mathrm{T} \mathop{}\!\mathrm{d}V},
	\label{eq: adjustment_coefficient}
\end{equation}

\noindent
where, ``$\forall$" stands for the~computational domain. Utilising this technique for welding simulations has already been reported in the~literature (see for instance, \cite{Meng_2016,Gao_2017,Wu_2019,Ebrahimi_2021}). Variations in energy absorption with deformations of the~melt-pool surface are, however, not taken into consideration in this method.

\paragraph{Case~3, heat source redistribution}
In this case, the~energy flux was redistributed over the~melt-pool surface assuming the~energy absorption is a~function of the~local surface orientation. Accordingly, the~local energy absorption is a~maximum where the~surface is perpendicular to the~laser ray and is a~minimum where the~surface is aligned with the~laser ray \cite{Arrizubieta_2017,Amara_2010}. This represents a~simplified absorptivity model based on the~Fresnel's equation \cite{Indhu_2018} and is expressed mathematically, assuming the~laser rays being parallel and in the~$z$-direction, as follows:

\begin{equation}
	S_\mathrm{T} =
	\begin{dcases}
		\frac{\eta\, Q}{\pi r_\mathrm{b}^2} \lVert \nabla\phi \rVert\frac{2\,\rho\, c_\mathrm{p}}{\left(\rho\, c_\mathrm{p}\right)_\mathrm{gas} + \left(\rho\, c_\mathrm{p}\right)_\mathrm{metal}} \left| \frac{\nabla \phi \left[z\right]}{\lVert \nabla \phi \rVert} \right|. & \text{if $r \le r_\mathrm{b}$} \\
		0. & \text{otherwise}
	\end{dcases}
	\label{eq: heat_source_case3}
\end{equation}

\noindent the~total energy input is not necessarily conserved in this method. 

\paragraph{Case~4, heat source redistribution and adjustment}
Utilising the~same technique introduced in Case~2, the~heat~source model in Case~3 was adjusted to guarantee that total energy input is conserved. Hence, the~heat~source model was defined as

\begin{equation}
	S_\mathrm{T} =
	\begin{dcases}
		\frac{\xi\, \eta\, Q}{\pi r_\mathrm{b}^2} \lVert \nabla\phi \rVert\frac{2\,\rho\, c_\mathrm{p}}{\left(\rho\, c_\mathrm{p}\right)_\mathrm{gas} + \left(\rho\, c_\mathrm{p}\right)_\mathrm{metal}} \left| \frac{\nabla \phi \left[z\right]}{\lVert \nabla \phi \rVert} \right|. & \text{if $r \le r_\mathrm{b}$} \\
		0. & \text{otherwise}
	\end{dcases}
	\label{eq: heat_source_case4}
\end{equation}

\paragraph{Case~5, Flat non-deformable free surface}
In this case, the~surface was assumed to remain flat. The~rmocapillary shear stresses caused by a~non-uniform temperature distribution on the~gas-metal interface were applied as a~boundary condition, hence modelling the~gas phase was not required. Consequently, the~total energy absorbed by the~surface was fixed in this method. The~boundary conditions are described in \cref{sec:boundary_conditions}. 

\Cref{tab:summary_of_cases} presents a~summary of the~cases considered in the~present work.

\begin{table*}[h] 
	\centering
	\caption{Summary of the~cases studied in the~present work and the~features included in the~model for each case.}
	\begin{tabular}{lccc}
		\hline
		Name   & Deformable free surface & Heat source adjustment & Heat source redistribution \\ \hline
		Case~1 &           Yes           &           No           &             No             \\
		Case~2 &           Yes           &          Yes           &             No             \\
		Case~3 &           Yes           &           No           &            Yes             \\
		Case~4 &           Yes           &          Yes           &            Yes             \\
		Case~5 &           No            &           No           &             No             \\ \hline
	\end{tabular} 
	\label{tab:summary_of_cases}
\end{table*}

\subsubsection{Boundary conditions}
\label{sec:boundary_conditions}

The bottom and lateral surfaces of the~metal part were modelled to be no slip walls, but in fact they remain solid during the~simulation time. At the~boundaries of the~gas layer above the~metal part, a~constant atmospheric pressure was applied (\textit{i.e.} $p = \SI{101.325}{\kilo\pascal}$) allowing air to flow in and out of the~domain. The~outer boundaries of the~computational domain were modelled as adiabatic since heat losses through these boundaries are negligible compared to the~laser power \cite{Kidess_2016_PhysFlu}.

For Case~5, in which no gas layer is modelled explicitly, a~thermocapillary shear-stress was applied as a~boundary condition in the~molten regions of the~top surface. In the~irradiated region on the~material top-surface, a~constant uniform heat flux was applied, while outside of this region the~surface was assumed to be adiabatic \cite{Pitscheneder_1996}. The~thermal and thermocapillary shear-stress boundary conditions were defined, respectively, as

\begin{equation}
	k \frac{\partial T}{\partial \vec{n}} = 
	\begin{dcases}
		\frac{\eta\, Q}{\pi r_\mathrm{b}^2}, & \text{if $r \le r_\mathrm{b}$} \\
		0, & \text{otherwise}
	\end{dcases}
	\label{eq: bc_heat_case5}
\end{equation}

\noindent
and

\begin{equation}
	-\mu \frac{\partial \vec{V_\mathrm{t}}}{\partial \vec{n}} = \frac{\mathop{}\!\mathrm{d}\sigma}{\mathop{}\!\mathrm{d}T} \frac{\partial T}{\partial \vec{\tau}},
	\label{eq: bc_shearstress_case5}
\end{equation}

\noindent
where, $\vec{V_\mathrm{t}}$ is the~tangential velocity vector, and $\vec{\tau}$ the~tangential vector to the~top surface.

\section{Numerical procedure}
\label{numerical_procedure}

The model was developed within the~framework of the~proprietary computational fluid dynamics solver ANSYS FLuent~\cite{Ansys}. The~laser heat source models and the~thermocapillary boundary conditions as well as the~surface-tension model were implemented through user-defined functions. After performing a~grid independence study ({results are presented in \cref{app:grid_study}}), a~grid containing $\SI{8.6e5}{}$ non-uniform hexahedral cells was utilised to discretise the~computational domain. Minimum cell spacing was $\SI{2e-5}{\meter}$ close to the~gas-metal interface and $\SI{3e-5}{\meter}$ in the~melt pool central region. Cell sizes gradually increase towards the~domain outer boundaries. The~computational domain employed in the~numerical simulations is shown in \cref{app:grid_study}. The~diffusion and convection terms in the~governing equations were discretised using the~central-differencing scheme with second order accuracy. For the~pressure interpolation, the~PRESTO~scheme~\cite{Patankar_1980} was used. Pressure and velocity fields were coupled employing the~PISO~scheme~\cite{Issa_1986}. An~explicit compressive VOF formulation was utilised for the~spatial discretisation of the~gas-metal interface advection \cite{Ubbink_1997}. The~transient advection terms were discretised using a~first order implicit scheme. To obtain a~Courant number $(\mathrm{Co} = \lVert\vec{V}\rVert\Delta t / \Delta x)$ less than $0.25$, with velocity magnitudes up to $\mathcal{O}(1)\, \SI{}{\meter\per\second}$, the~time-step size was set to $10^{-5}\, \SI{}{\second}$. Each simulation was executed in parallel on $40$ cores (Intel~Xeon E5-2630~v4) of a~high-performance computing cluster. Scaled residuals of the~energy, momentum and continuity equations of less than $10^{-10}$, $10^{-8}$ and $10^{-7}$ respectively, were defined as convergence~criteria.

\section{Results and Discussion}
\label{sec:results}

\subsection{Model validation and solver verification}
\label{sec:validation}

\begin{figure}[H] 
	\centering
	\includegraphics[width=0.75\linewidth]{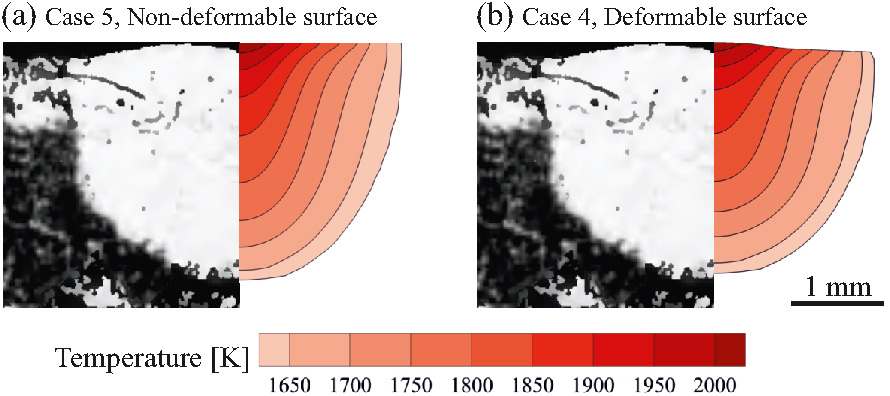}
	\caption{Comparison of the~numerically predicted melt-pool shape after $\SI{5}{\second}$ of heating with the~corresponding experimentally measured post-solidification melt-pool shape reported by Pitscheneder \textit{et~al.}~\cite{Pitscheneder_1996}. Laser power was set to $\SI{3850}{\watt}$ and the~material contains $\SI{150}{ppm}$ sulphur. Numerical predictions obtained from the~present model assuming (a)~non-deformable (Case~5) and (b) deformable gas-metal interface (Case~4).}
	\label{fig:validation}
\end{figure}

To verify the~reliability and accuracy of the~present numerical simulations, the~melt-pool shapes obtained from the~present simulations, for the~problem introduced in \cref{sec:phys_model}, are compared to experimental observations reported by \mbox{Pitscheneder \textit{et~al.}~\cite{Pitscheneder_1996}}. \Cref{fig:validation} shows a~comparison between the~melt-pool shape obtained from the~numerical simulation after $\SI{5}{\second}$ of heating and the~post-solidification experimental observation, which indicates a~reasonable agreement. The~maximum absolute deviations between the~present numerical predictions and experimental data for the~melt-pool width and depth is less than $5\%$ and $2\%$, respectively. However, it should be noted that the~thermal conductivity and viscosity of the~liquid material were artificially increased in the~simulations by a~factor $\mathcal{F}=7$ with respect to their reported experimental values, as suggested by \mbox{Pitscheneder \textit{et~al.}~\cite{Pitscheneder_1996}}. This is known as employing an~``enhancement factor", mainly to achieve agreement between numerical and experimental results \cite{Saldi_2013}. Independent studies conducted by~\mbox{Saldi \textit{et~al.}~\cite{Saldi_2013}} and Ehlen \textit{et~al.}~\cite{Ehlen_2003} revealed that without using such an~enhancement factor, the~fluid flow structure and the~melt-pool shape in simulations differ drastically from experimental observations. The~use of an~enhancement factor is often justified by the~possible occurrence of turbulence and its influence on heat and momentum transfer in the~melt~pool, which is assumed to be uniform in the~melt pool. However, the~high-fidelity numerical simulations conducted by \mbox{Kidess \textit{et~al.}~\cite{Kidess_2016,Kidess_2016_PhysFlu}} on a~melt pool with a~flat non-deformable melt-pool surface revealed that turbulent enhancement is strongly non-uniform in the~melt pool, resulting in an~$\omega$-shaped melt pool that differs notably from the~results of simulations assuming uniform transport enhancement. We extended this study and performed a~high-fidelity simulation based on the~large eddy simulation (LES) turbulence model and took the~effects of surface deformations into consideration \cite{Ebrahimi_2019_2} and showed that for the~conduction-mode laser melting problem we considered, the~influence of surface oscillations on melt pool behaviour is larger compared to the~effects of turbulent flow in the~melt pool; nevertheless the~results do not agree with experiments without using some enhancement. These observations along with previous studies \cite{Saldi_2013,Ehlen_2003,Mundra_1992} suggest that the~published weld pool models lack the~inclusion of significant physics. The~neglect in simulations of relevant physics such as chemical reactions and unsteady oxygen absorption by the~melt-pool surface, non-uniform unsteady surfactant distribution over the~melt-pool surface, re-solidification, free surface evolution and three-dimensionality of the~fluid flow field has been postulated as reasons why such an~\textit{ad hoc} and unphysical enhancement factor is needed to obtain agreement with experimental data \cite{Do_Quang_2008,Winkle_2000,Tsai_1989}. However, the~inclusion of such factors will increase the~model complexity and computational costs. Further investigations are essential to realising the~problem with sufficient details. In \cref{sec:the_effects_of_enhancement_factor}, the~effects of the~value used for the~enhancement factor $\mathcal{F}$ on the~thermal and fluid flow fields and melt-pool oscillatory behaviour are discussed in more detail.

\begin{figure*}[htbp]
	\centering
	\includegraphics[width=0.75\linewidth]{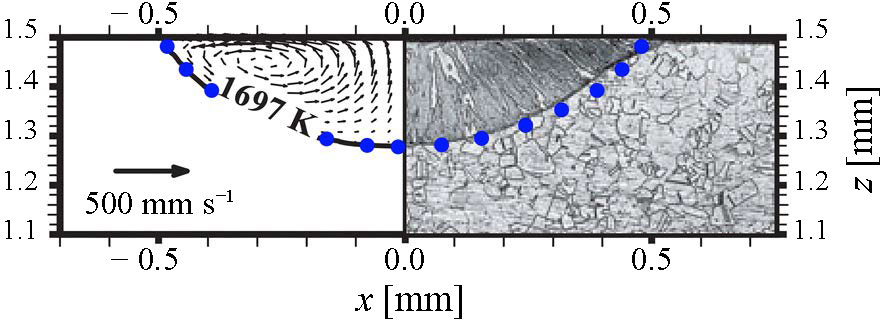}
	\caption{Comparison of the~spot melt-pool shape obtained from the~present numerical simulation with the~corresponding experimentally observed (the right side of the~figure) and numerically predicted (the left side of the~figure) post-solidification melt-pool shape reported by He \textit{et~al.}~\cite{He_2003}. Laser power was set to $\SI{1967}{\watt}$ and the~beam radius was $\SI{570}{\micro\meter}$. Blue circles show the~melt-pool shape predicted using the~present model. The~$\SI{500}{\milli\meter\per\second}$ reference vector is provided for scaling the~velocity field shown in the~melt pool.}
	\label{fig:validation-exp-2}
\end{figure*}

The experimental and numerical data reported by He \textit{et~al.}~\cite{He_2003}, who investigated unsteady heat and fluid flow in the~melt pool during conduction-mode laser spot welding of stainless steel (SS304) plates, were considered to validate the~reliability of the~present model. The~melt pool shape obtained from the~present model assuming a~non-deformable free surface (Case 5) was compared with experimental and numerical data reported by He \textit{et~al.}~\cite{He_2003} and the~results are shown in \cref{fig:validation-exp-2}. In this problem, the~laser power was set to $\SI{1967}{\watt}$, the~beam radius was $\SI{570}{\micro\meter}$ and laser pulse duration was $\SI{3}{\milli\second}$. The~thermal conductivity and viscosity of the~molten material were artificially increased in the~simulations by a~factor $\mathcal{F} = 17$ and $\mathcal{F} = 11$, respectively, as suggested by He \textit{et~al.}~\cite{He_2003}. The~maximum absolute deviations between the~present numerical predictions and experimental data for the~melt-pool width and depth is less than $8\%$ and $2.5\%$, respectively. The~results obtained from the~present model are indeed in reasonable agreement with the~reference data, indicating the~validity of the~present model in reproducing the~results reported in literature.

\begin{figure}[H]
	\centering
	\includegraphics[width=1.00\linewidth]{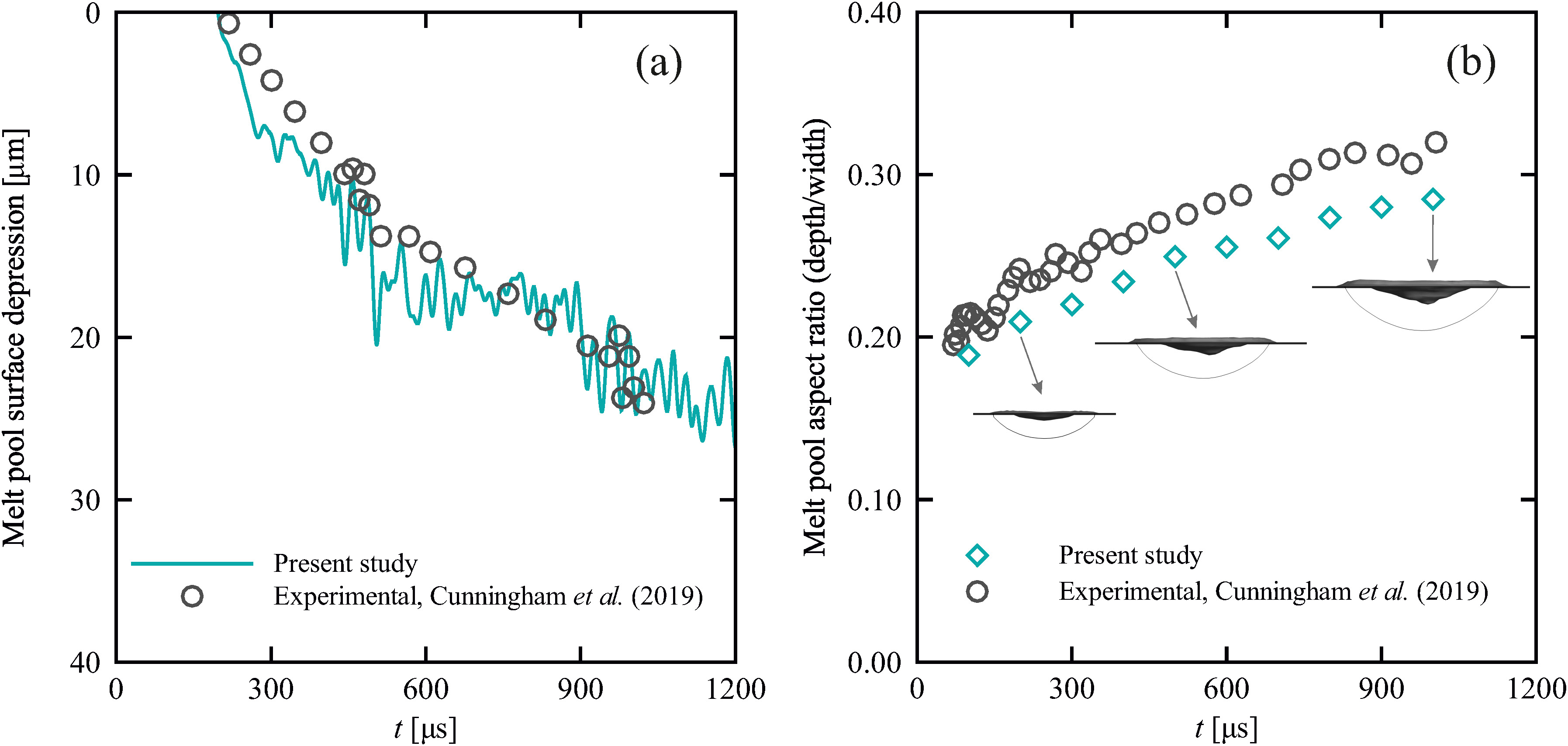}
	\caption{Comparison of (a) the~melt-pool surface depression and (b) the~melt pool aspect ratio (depth to width ratio) obtained from the~present numerical simulation with the~corresponding experimental observations of Cunningham \textit{et~al.}~\cite{Cunningham_2019}. Laser power was set to $\SI{156}{\watt}$ and the~beam radius was $\SI{70}{\micro\meter}$. Snapshots of the~melt pool and its deformed surface are shown in the~right subfigure for three different time instances.}
	\label{fig:validation-exp}
\end{figure}

{
	The reliability of the~present model was also investigated by comparing the~numerically predicted melt-pool size and surface deformations with experimental observations of Cunningham \textit{et~al.}~\cite{Cunningham_2019}. The~evolutions of the~melt pool shape and its surface depression under stationary laser melting of a~Ti-6Al-4V plate under conduction mode were studied. In this problem, the~laser power is $\SI{156}{\watt}$ and has a~Gaussian distribution and the~radius of the~laser spot is $\SI{7e-5}{\meter}$. The~thermophysical properties of Ti-6Al-4V suggested by~\mbox{Sharma \textit{et~al.}~\cite{Sharma_2018}} were employed for calculations. Since the~melt pool surface temperature reaches the~boiling point, the~effects of recoil pressure $p_\mathrm{r}$ and evaporation heat loss $q_\mathrm{e}$ were included in the~model. The~recoil pressure~\cite{Lee_2002} and evaporation heat loss~\cite{von_Allmen_1995} were determined respectively as follows:}

\begin{equation}
	p_\mathrm{r} =  0.54 \cdot P_0 \exp\left(\frac{L_\mathrm{v} \, \mathrm{M} \, \left(T - T_\mathrm{v}\right)}{\mathrm{R} \, T \, T_\mathrm{v}}\right),
	\label{eq:recoil_pressure}
\end{equation}

\begin{equation}
	q_\mathrm{e} =  -0.82 \frac{L_\mathrm{v} \, \mathrm{M}}{\sqrt{2 \pi \mathrm{M} \, \mathrm{R} \, T}} P_0 \exp\left(\frac{L_\mathrm{v} \, \mathrm{M} \, \left(T - T_\mathrm{v}\right)}{\mathrm{R} \, T \, T_\mathrm{v}}\right),
	\label{eq:evaporation_heat_loss}
\end{equation}

\noindent
where, $P_0$ is the~ambient pressure, $L_\mathrm{v}$ the~latent heat of vaporisation, $\mathrm{M}$ the~molar mass, $\mathrm{R}$ the~universal gas constant, $T$ the~temperature and $T_\mathrm{v} = \SI{3315}{\kelvin}$ the~evaporation temperature. The~results obtained from the~present numerical simulations using the~heat source model introduced for Case 4 are compared with those reported by Cunningham \textit{et~al.}~\cite{Cunningham_2019} in \cref{fig:validation-exp}, which show a~reasonable agreement. The~maximum absolute deviations between the~present numerical predictions and experimental data for the~melt-pool aspect ratio is less than $12\%$.

{Particular attention was also paid to the~effect of spurious currents on numerical predictions of molten metal flow in melt pools with free surface deformations. Spurious currents are an~unavoidable numerical artefact generating unphysical parasitic velocity fields that do not vanish with grid refinement when using the~continuum surface force technique for modelling surface tension effects in the~VOF method. This~problem has been thoroughly investigated by \mbox{Mukherjee \textit{et~al.}~\cite{Mukherjee_2018}} and the~results of their study show that the~Reynolds number based on the~time-averaged maximum spurious velocity is $\mathcal{O}(10)$. The~results of the~present simulations show that the~flow Reynolds number is $\mathcal{O}(10^{3})$ that is at least two orders of magnitude larger than that induced by spurious currents. Furthermore, to suppress fluid velocities in the~solid regions a~momentum sink term is defined based on the~enthalpy porosity technique that further weakens the~spurious currents in solid regions. The~refore, the~influence of spurious currents on numerical predictions is considered to be negligible in the~present study.}

\subsection{The influence of heat source adjustment}
\label{sec:influence_of_heat_source_adjustment}

When accounting for surface deformations, the~volume-of-fluid (VOF) method developed by Hirt and Nichols~\cite{Hirt_1981}, based on a~Eulerian formulation, is the~most common method for modelling the~melt~pool behaviour. In this diffuse boundary method, the~interfacial forces and the~energy fluxes applied on the~melt-pool surface are treated as volumetric source terms in the~surface region, instead of imposing them as boundary conditions. In this approach, however, the~fact that surface deformations lead to temporal and spatial variations of the~free~surface boundary conditions, as remarked by Meng \textit{et~al.}~\cite{Meng_2016} and Wu \textit{et~al.}~\cite{Wu_2019}, is often neglected. The~results reported by \mbox{Choo \textit{et~al.}~\cite{Choo_1990}} suggest that variations in power-density distribution and changes in free-surface profile can affect molten metal flow in melt pools and its stability. In this section, the~influence of applying each of the~five different approaches to model the~laser heat source (described as Cases 1--5 in \cref{sec:laser_heat_source}) on the~melt pool behaviour is discussed. In these simulations, no~enhancement~factor, as introduced in the~previous section, was employed since it is \textit{ad hoc} and has little justification in physical reality.

The energy flux distribution over the~melt-pool surface determines the~spatial temperature distribution over the~melt-pool surface and its temporal variations. Due to the~temporal changes in the~melt-pool surface shape, the~total energy absorbed by the~material in Case~1 varies between $\SI{491.4}{\watt}$ and $\SI{538.1}{\watt}$, with a~median value of $\SI{515.7}{\watt}$, which differs from the~total energy supplied by the~laser $\eta Q = \SI{500.50}{\watt}$. This~issue is resolved by utilising a~dynamic adjustment coefficient in Case~2, while the~relative energy flux distribution remains unchanged. However, the~results reported by Courtois \textit{et~al.}~\cite{Courtois_2016} and \mbox{Bergstr{\"o}m \textit{et~al.}~\cite{Bergstr_m_2007}} showed that the~energy flux distribution varies with surface deformation during a~laser melting process. When surface deformations are too small to cause multiple reflections, redistributing the~energy flux over the~free surface results in a~better input energy conservation as obtained in Case~3. The~total energy absorbed by the~material in Case~3 ranges between $\SI{482.6}{\watt}$ and $\SI{532.3}{\watt}$, with a~median value of $\SI{503.8}{\watt}$. By further introducing an~adjustment coefficient (\cref{eq: adjustment_coefficient}), the~absorbed energy can be made to exactly match the~supplied laser power $\eta Q$. Finally, for the~most simple approach without any surface deformations (\textit{i.e.}~Case~5), the~total absorbed energy exactly matches the~supplied laser power and remains unchanged in time.

\begin{figure}[H] 
	\centering
	\includegraphics[width=1.00\linewidth]{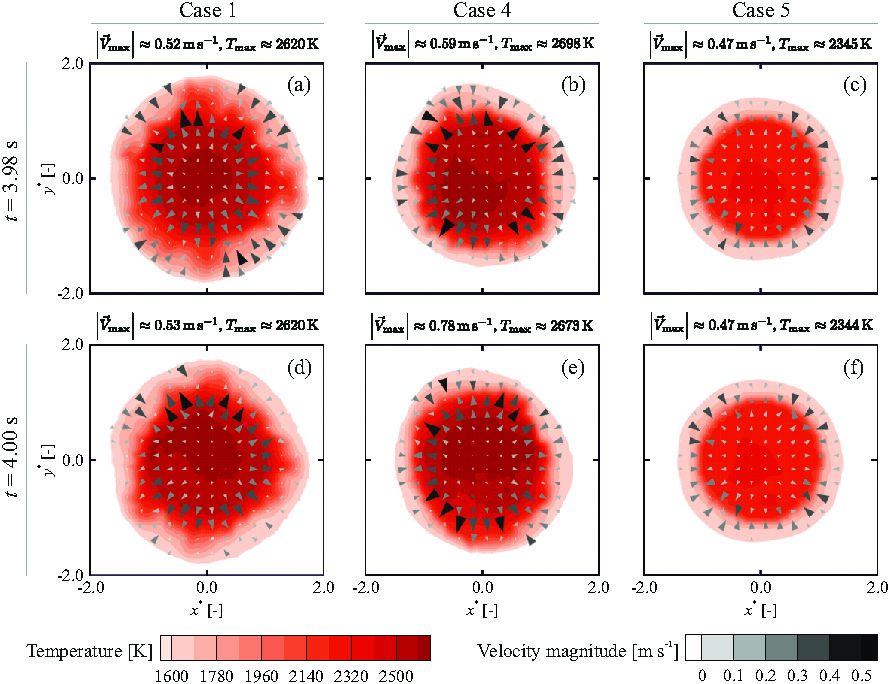}
	\caption{Contours of temperature and the~velocity vectors on the~melt pool free surface at two different time instances for Case~1, 4 and 5. Coordinates are non-dimensionalised using the~laser-beam radius $r_\mathrm{b}$ as the~characteristic length scale.}
	\label{fig:temperature_velocity}
\end{figure}

The variations of temperature distribution over the~melt-pool surface over time determine the~thermocapillary driven flow pattern and melt-pool shape, as shown in \cref{fig:temperature_velocity}. An~unsteady, asymmetric, outwardly directed fluid flow emanating from the~pool centre is found for all 5 cases. Flow accelerates towards the~pool rim, where it meets inwardly directed fluid flow from the~rim, due to the~change of sign of $\left(\partial \sigma / \partial T\right)$ at a~certain pool temperature (see \cref{fig:surface_tension_variations}). Close to the~region where these two flows meet, velocity magnitudes are high due to the~large thermocapillary stresses generated by the~steep temperature gradients, in~contrast to the~low flow velocities in the~central region of the~pool surface. Interactions between these two opposing flows, and their interactions with the~fusion boundary, cause the~fluid flow inside the~melt pool to be asymmetric and unstable \mbox{\cite{Kidess_2016_PhysFlu}}, leading to a~distorted melt-pool shape.

Melt-pool surface temperatures are roughly $7\%$--$16\%$ higher when surface deformations are taken into consideration (Cases 1--4), compared to flat surface simulations (Case~5). Temperature gradients over the~melt-pool surface are also on average larger in Case~1--4 (with a~deformable surface) and particularly in Case~2--4 (with heat source adjustment and/or redistribution), compared to Case~5 (with a~non-deformable surface). Consequently, local fluid velocities caused by thermocapillary stresses are almost $25\%$ higher for Case~1--4, compared to those of Case~5, and are of the~order of $\SI{0.6}{\meter\per\second}$, in reasonable agreement with experimental measurements performed by Aucott \textit{et~al.}~\cite{Aucott_2018} and estimated values from the~scaling analyses reported by Oreper and Szekely~\cite{Oreper_1984}, Rivas and Ostrach~\cite{Rivas_1992} and \mbox{Chakraborty and Chakraborty~\cite{Chakraborty_2007}}.

Free surface deformations have a~destabilising effect on the~melt pool due to the~augmentation of temperature gradients over the~melt-pool surface, leading to higher thermocapillary stresses \cite{Davis_1980}. Additionally, the~stagnation region, where the~sign of $\left(\partial \sigma / \partial T\right)$ and thus thermocapillary flow direction change, is sensitive to small spatial disturbances and further enhances melt pool instabilities. Furthermore, variations in magnitude and direction of velocities can cause rotational and pulsating fluid motions, leading to cross-cellular flow patterns with a~stochastic behaviour in the~melt pool \cite{Czerner_2005,Zhao_2010}.

All such flow instabilities reinforce unsteady energy transport from the~melt pool to the~surrounding solid material, resulting in continuous melting and re-solidification of the~material close to the~solid-liquid boundary. This complex interplay leads to the~melt-pool surface width for Case~5 to be $23\%$ smaller than that for Case~1, and $10\%$ smaller than for Case~2--4. The~higher amount of heat absorbed by the~deformed free-surface in Case~1, and the~enhanced convective heat transfer in Case~1--4, are the~main causes for the~observed pool size differences.

\begin{figure}[H] 
	\centering
	\includegraphics[width=0.75\linewidth]{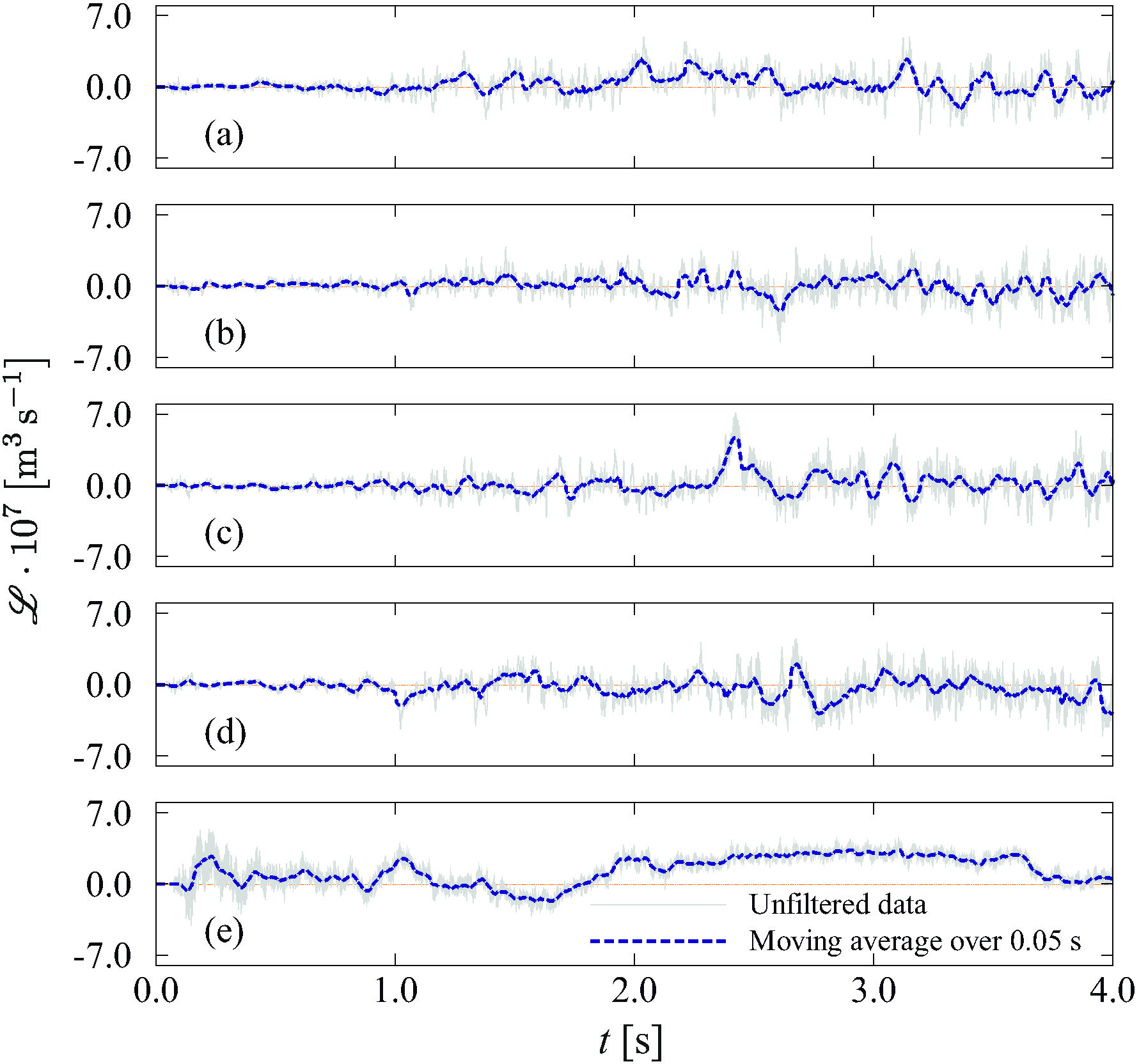}
	\caption{Variations of the~averaged angular momentum over the~melt-pool surface over a~time period of $\SI{4}{\second}$ for (a)~Case~1, (b)~Case~2, (c)~Case~3, (d)~Case~4 and (e)~Case~5. Unfiltered data: grey lines, Moving average of the~data over \SI{0.05}{\second}: dashed blue lines. $\mathscr{L} > 0$ indicates a~clockwise rotation.}
	\label{fig:angular_momentum}
\end{figure}

To investigate the~rotational fluid motion, the~spatially-averaged angular momentum over the~melt-pool surface ($\mathscr{L}$) about the~$z$-axis (\textit{i.e.} the~axis of rotation) with respect to the~origin $\mathscr{O}(0, 0, 0)$ is calculated as follows:

\begin{equation}
	\mathscr{L}\left(t\right) = \frac{1}{A}\iint \limits_{\mathrm{free\ surface}} \hat{z} \cdot \left(\vec{r} \times \vec{V}\right) \mathop{}\!\mathrm{d}A,
	\label{eq: angular_momentum}
\end{equation}

\noindent
where, $\hat{z}$ is a~unit vector in the~z-direction, $\vec{r}$ the~position vector, $\vec{V}$ velocity vector and $A$ the~area of gas-metal interface. \Cref{fig:angular_momentum} shows the~temporal variations of angular momentum over the~melt-pool surface as a~function of time. Positive values of the~angular momentum in \cref{fig:angular_momentum} show a~clockwise flow rotation, and vice versa. Kidess \textit{et~al.}~\cite{Kidess_2016_PhysFlu} applied a~similar approach to investigate rotational fluid motion over a~flat non-deformable melt-pool surface. Continuous fluctuations in the~sign of angular momentum indicate an~oscillatory rotational fluid motion in the~pool that is indeed self-excited. In Case~1--4, flow pulsations start to take place after roughly $\SI{1}{\second}$ and continue thereafter. This is not however valid for Case~5, where flow pulsations start to occur already after about $\SI{0.1}{\second}$, and a~clockwise rotation is established after $\SI{1.8}{\second}$, which lasts for roughly two seconds. These instabilities in the~flow field are attributed to a~large extent to the~interactions between the~two opposing flows meeting at the~melt-pool surface  \cite{Kidess_2016_PhysFlu,Kidess_2016}, and to a~lesser extent the~effects of hydrothermal waves \cite{Kuhlmann_2010,Davis_1987}.

\begin{figure}[H] 
	\centering
	\includegraphics[width=1.00\linewidth]{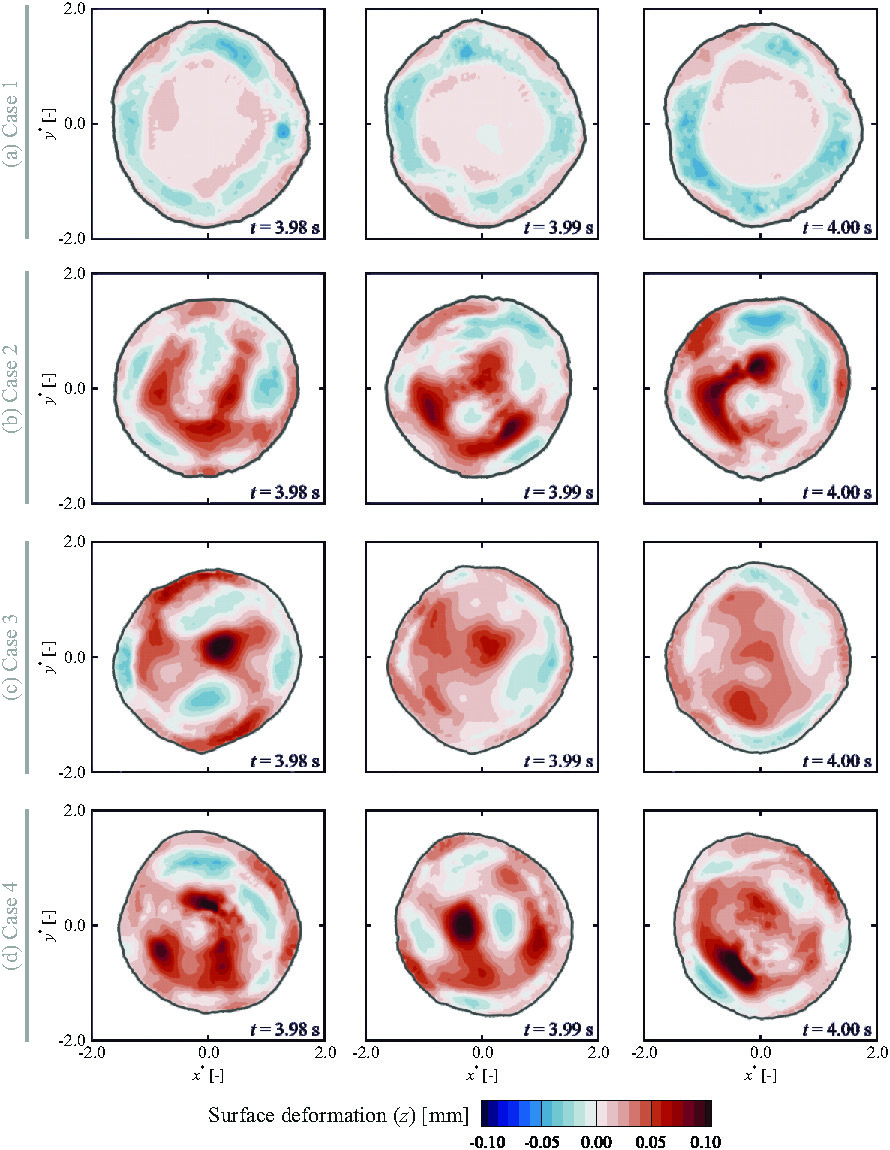}
	\caption{Contours of melt pool free surface deformation at three different time instances for (a)~Case~1, (b)~Case~2, (c)~Case~3, and (d)~Case~4. Coordinates are non-dimensionalised using the~laser-beam radius $r_\mathrm{b}$ as the~characteristic length scale. Positive and negative values of \textit{z} indicate surface depression and elevation, respectively.}
	\label{fig:surface_deformations}
\end{figure}

Free surface deformations evolve rapidly because of the~thermocapillary stresses acting on the~melt-pool surface and are shown in \cref{fig:surface_deformations} at three different time instances. Free surface deformations are smaller and less intermittent in Case~1 compared with Case~2--4. This is due to the~smaller thermocapillary stresses and the~wider stagnation region, which makes the~surface flow field less sensitive to spatial disturbances. Additionally, the~Capillary number ($\mathrm{Ca} = \mu \lVert\vec{V}\rVert / \sigma$), which represents the~ratio between viscous and surface-tension forces and which is $\mathcal{O}(10^{-3})$ for all the~cases studied in the~present work, appears to be larger in Case~2--4 ($\mathrm{Ca} \approx \SI{2.5e-3}{}$) compared with Case~1 ($\mathrm{Ca} \approx \SI{1.5e-3}{}$), particularly in the~central region of the~melt-pool surface. Redistributing the~energy flux based on the~free surface profile (Case~3 and~4) disturbs the~temperature field locally and thus the~velocity distribution over the~melt-pool surface. 

Fluid flow may influence the~energy transport and associated phase change significantly during laser melting \cite{DebRoy_1995}, which can be assessed through the~P{\'e}clet number that is the~ratio between advective and diffusive heat transport, defined as follows:

\begin{equation}
	\mathrm{Pe} = \frac{\mathscr{D} \lVert\vec{V}\rVert \rho c_\mathrm{p}}{k},
	\label{eq: peclet}
\end{equation}

\noindent
where, $\mathscr{D}$ is a~characteristic length scale, here chosen to be $r_\mathrm{b}$. The~value of the~P{\'e}clet number is much larger than one ($\mathrm{Pe} = \mathcal{O}(10^2)$), which indicates a~large contribution of advection to the~total energy transport, compared with diffusion.

\begin{figure}[H] 
	\centering
	\includegraphics[width=1.00\linewidth]{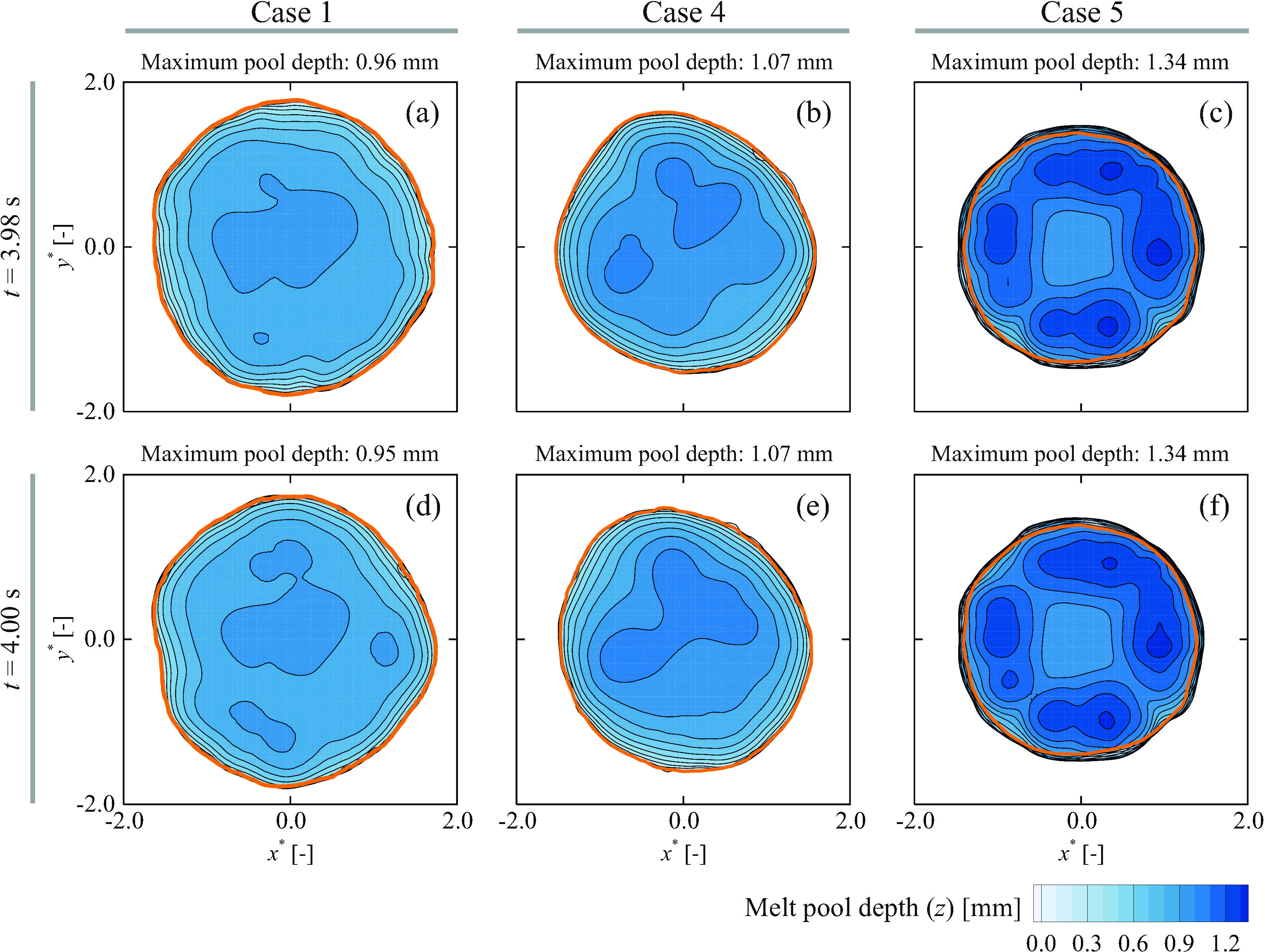}
	\caption{Contours of melt-pool depth at $t =$ 3.98 and $\SI{4.00}{\second}$ for Case~1, 4 and 5. Coordinates are non-dimensionalised using the~laser-beam radius $r_\mathrm{b}$ as the~characteristic length scale. The~orange circular line shows the~melt pool boundary at its top surface.}
	\label{fig:pool_depth}
\end{figure}

The predicted melt-pool depth is shown in \cref{fig:pool_depth} at two different time instances. When surface deformations are taken into account (Case 1--4), the~predicted melt-pool shape is shallow, with its maximum depth located in the~central region and its largest width at the~surface. When the~surface is assumed to be non-deformable (Case 5), on the~other hand, the~depth profile of the~melt-pool resembles a~doughnut-shaped, with its maximum width located below the~gas-metal interface. The~predicted fusion boundaries of all the~cases studied in the~present work are asymmetric and unstable. Flow instabilities in the~melt pool increase when surface deformations are taken into account, which enhance convection in the~pool resulting in a~smooth melt-pool depth profile. Variations in the~melt-pool shape for Case~5 are less conspicuous compared to those for Case~1--4 because of the~smaller fluctuations in the~fluid flow field.

\begin{figure}[H] 
	\centering
	\includegraphics[width=1.00\linewidth]{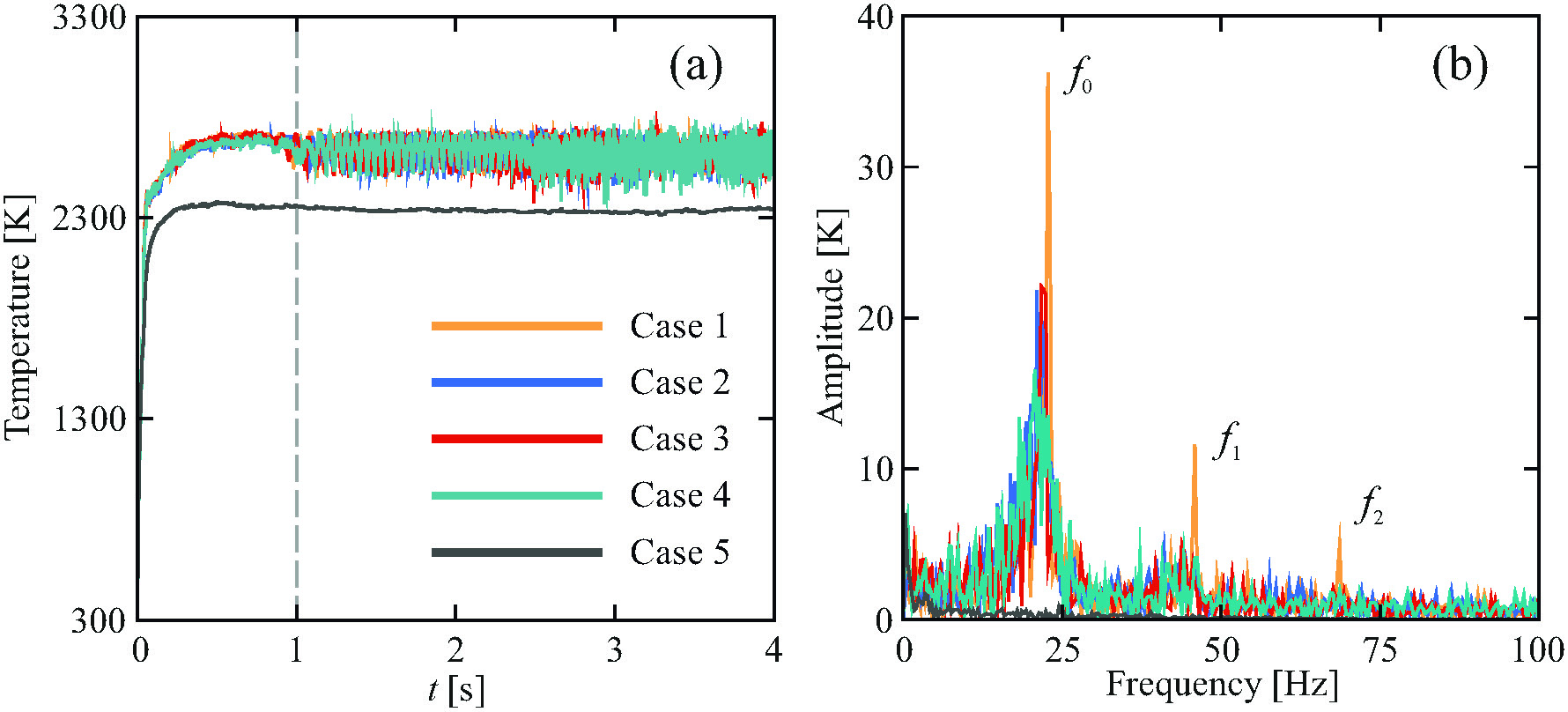}
	\caption{(a) Temperature signals recorded from the~monitoring point $p_4$ located at $\left(x^*, y^*\right) = \left(0, 0\right)$ on the~melt-pool surface and (b)~the~corresponding frequency spectra. Temperature signals in the~period of $1$ to $\SI{4}{\second}$ are employed for FFT analysis.}
	\label{fig:temperature_pool_centre}
\end{figure}

To understand the~three-dimensional oscillatory flow behaviour, three monitoring points were selected over the~melt-pool surface in different non-azimuthal directions $p_1\left(x^*, y^*\right) = \left(5/7,0\right)$, $p_2\left(x^*, y^*\right) = \left(5/7,5/7\right)$ and $p_3\left(x^*, y^*\right) = \left(0,5/7\right)$ in addition to a~point $p_4\left(x^*, y^*\right) = \left(0,0\right)$ on the~melt-pool surface, in which $x^*$ and $y^*$ are non-dimensionalized with the~laser beam radius $r_b$. Temperature signals, recorded from the~monitoring point $p_4$, and the~corresponding ``Fast~Fourier~Transform"~(FFT) \cite{Temperton_1985} frequency spectra are shown in \cref{fig:temperature_pool_centre}. In the~presence of surface deformations (Case~1--4), self-excited temperature fluctuations, initiated by numerical noise, grow and reach a~quasi-steady state after about \SI{1}{\second}. Without surface deformations (Case~5), on the~other hand, the~amplitudes of temperature fluctuations remain very small.	In Case~1, where surface deformations are taken into account, the~temperature fluctuations with large amplitudes have a~fundamental frequency of $f_\mathrm{0} \approx \SI{22}{\hertz}$, with harmonics at $f_1 \approx 2 f_\mathrm{0}$ and $f_2 \approx 3 f_\mathrm{0}$. Irregular patterns appear in the~spectrum of temperature fluctuations for Case~2--4, showing a~highly unsteady behaviour that results from the~enhancement of thermocapillary stresses and the~variations in energy flux distribution.

\begin{figure}[H] 
	\centering
	\includegraphics[width=0.75\linewidth]{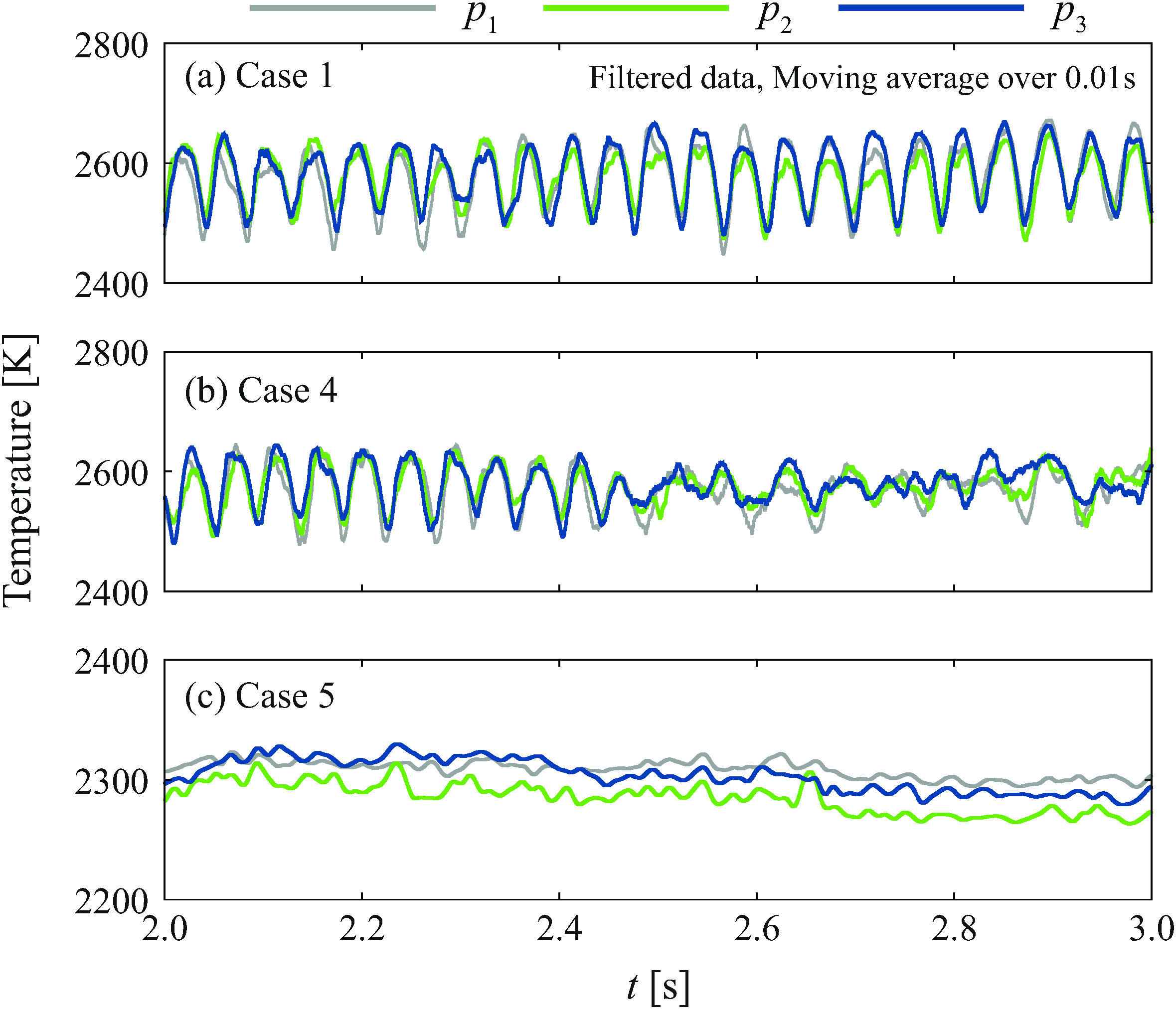}
	\caption{Temperature signals recorded from three monitoring points at the~melt pool free surface placed at a~radius of $\SI{1}{\milli\meter}$ and along different azimuthal directions ($p_1\left(x^*, y^*\right) = \left(5/7,0\right)$, $p_2\left(x^*, y^*\right) = \left(5/7,5/7\right)$ and $p_3\left(x^*, y^*\right) = \left(0,5/7\right)$). (a)~Case~1, (b)~Case~4, and (c)~Case~5. Signals are smoothed using a~moving averaging window of $\SI{0.01}{\second}$.}
	\label{fig:temperature_monitoring_point}
\end{figure}

Temperature signals recorded from monitoring points $p_1$, $p_2$ and $p_3$ in the~time interval of $2$--$\SI{3}{\second}$ are shown in \cref{fig:temperature_monitoring_point}. The~data for Case~1 indicate that the~thermal and fluid flow fields are dominated by a~pulsating behaviour. This is also valid for Case~4 up to roughly $\SI{2.5}{\second}$; however, after $\SI{2.5}{\second}$ modulation of temperature fluctuations takes place, which is probably due to changes in temperature gradients, melt pool size and the~complex interactions between vortices generated inside the~melt pool. Similar behaviour was observed for Case~2 and 3 (not shown here). The~amplitudes of temperature fluctuations appear to be smaller for Case~5 compared to the~other cases and show an~irregular behaviour, revealing complex unsteady flow in the~melt pool.

\subsection{The effects of employing temperature-dependent material properties}
\label{sec:the_effects_of_temperature_dependent_properties}

\begin{figure}[H] 
	\centering
	\includegraphics[width=0.75\linewidth]{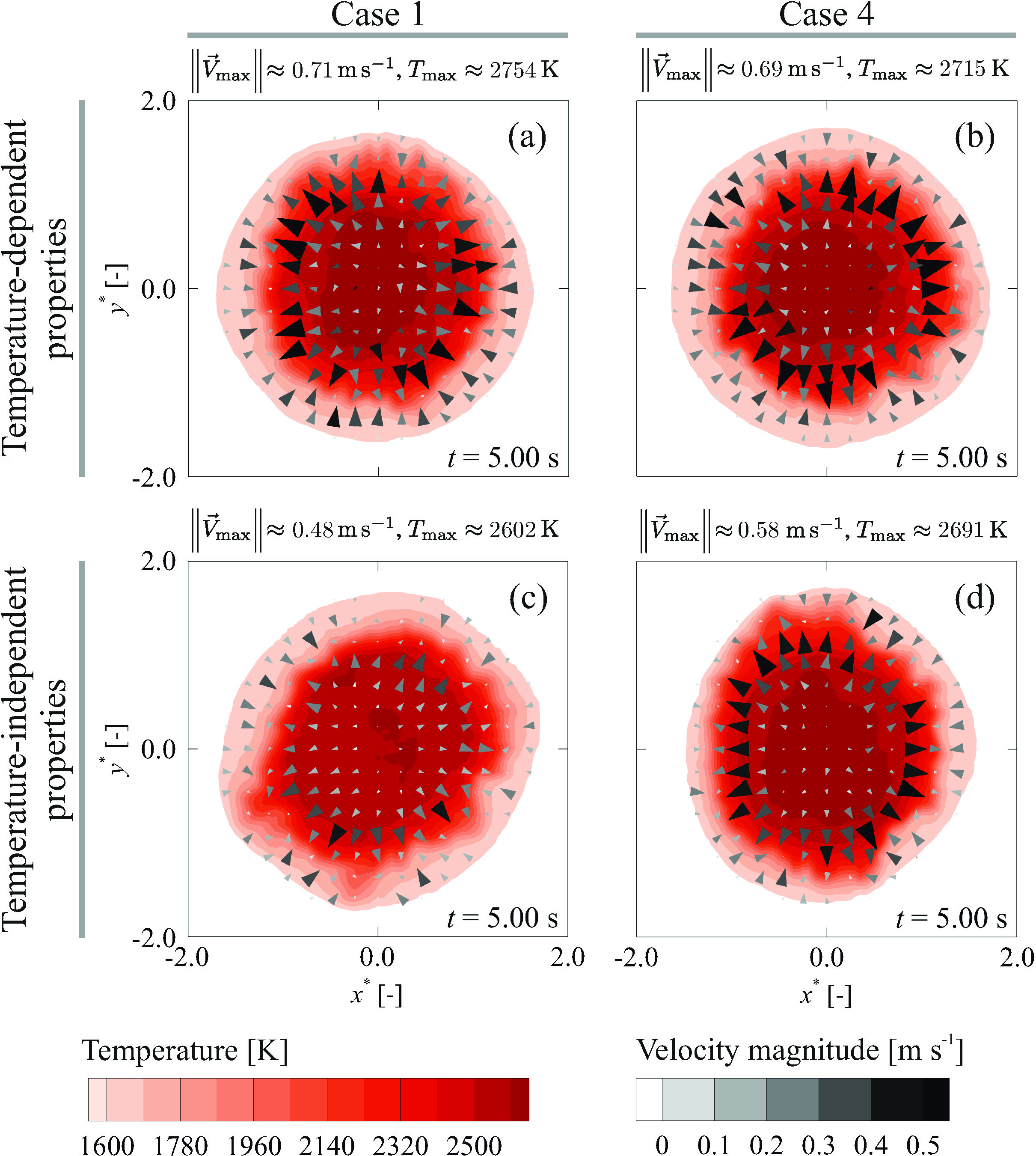}
	\caption{Contours of temperature and the~velocity vectors on the~melt pool free surface after $\SI{5}{\second}$ of heating predicted using the~heat source models Case~1~and~4 with temperature-dependent (\cref{tab:temperature-dependent_material_properties}) and temperature-independent (\cref{tab:material_properties}) material properties. Coordinates are non-dimensionalised using the~laser-beam radius $r_\mathrm{b}$ as the~characteristic length scale.}
	\label{fig:temperature_velocity_s705}
\end{figure}

In this section, the~effects of employing temperature-dependent properties on numerical predictions of thermal and flow fields as well as the~melt-pool shape are investigated using two different heat source models (Case~1~and~4). Temperature-dependent properties for the~metallic alloy considered in the~present study~(S705) are given in \cref{tab:temperature-dependent_material_properties}, where the~values are estimated by analogy with the~values for iron-based alloys \cite{Mills_2002}. Temperature distribution over the~melt-pool surface and free-surface flow after $\SI{5}{\second}$ of heating are shown in \cref{fig:temperature_velocity_s705} for both temperature-dependent and temperature-independent thermophysical properties. Melt-pool surface temperatures predicted using temperature-independent properties are about $1$--$6\%$ lower than those predicted using temperature-dependent properties. This is mainly attributed to the~changes in the~thermal diffusivity $\left(\alpha = k / \left(\rho c_\mathrm{p}\right)\right)$ of the~molten material. The~thermal conductivity of the~molten material changes from $\SI{21.8}{\watt\per\meter\per\kelvin}$ at $T = \SI{1620}{\kelvin}$ to $\SI{33.4}{\watt\per\meter\per\kelvin}$ at $T = \SI{2700}{\kelvin}$. Hence, the~average thermal diffusivity of the~molten material obtained from a~temperature-dependent model is about $10\%$ lower than that estimated using temperature-independent properties. The~fluid velocities are roughly $15$--$32\%$ lower when temperature-independent properties are employed, compared to those predicted using temperature-dependent properties, which is due to the~decrease in the~viscosity of the~molten material at elevated temperatures. With an~increase in temperature, the~thermal conductivity of the~molten metal increases and the~viscosity of the~molten metal decreases, resulting in a~reduction of momentum diffusivity and enhancement of thermal diffusivity and thus reduction of the~Prandtl number $\left(\mathrm{Pr} = c_\mathrm{p} \, \mu / k\right)$. Although the~heat source models affect the~thermal and flow fields in the~melt pool, it is found that the~numerical predictions are less sensitive to the~heat source models when temperature-dependent properties are employed for the~cases studied in the~present work. However, it should be noted that for the~cases where surface deformations are larger, the~effects of heat source adjustment on numerical predictions become critical, as discussed in \cref{sec:influence_of_heat_source_adjustment}.

\begin{table}[H] 
	\centering
	\caption{Temperature-dependent thermophysical properties of the~Fe-S alloy used in the~present study. Values are estimated by analogy with the~values for iron-based alloys \cite{Mills_2002}.}
	\begin{tabular}{lll}
		\hline
		Property                              & Fe-S alloy                                                                  & Unit                                \\ \hline
		Density $\rho$                        & \SI{8100}                                                                   & \si{\kilogram\per\meter\cubed}      \\
		Specific heat capacity $c_\mathrm{p}$ & \SI{627} (solid phase)                                                      & \si{\joule\per\kilogram\per\kelvin} \\
		& \SI{723.14} (liquid phase)                                                  &                                     \\
		Thermal conductivity $k$              & $\SI{8.8521} + \SI{0.0114}\cdot T$ (solid phase)                            & \si{\watt\per\meter\per\kelvin}     \\
		& $\SI{4.5102} + \SI{0.0107}\cdot T$ (liquid phase)                           &                                     \\
		Viscosity $\mu$                       & $\SI{0.0659} - \SI{7e-5}\cdot T + \SI{3e-8}\cdot T^2 - \SI{3e-12}\cdot T^3$ & \si{\kilogram\per\meter\per\second} \\
		Latent heat of fusion $L_\mathrm{f}$  & \SI{250800}                                                                 & \si{\joule\per\kilogram}            \\
		Thermal expansion coefficient $\beta$ & \SI{2e-6}                                                                   & \si{\per\kelvin}                    \\
		Liquidus temperature $T_\mathrm{l}$   & \SI{1620}                                                                   & \si{\kelvin}                        \\
		Solidus temperature $T_\mathrm{s}$    & \SI{1610}                                                                   & \si{\kelvin}                        \\ \hline
	\end{tabular} 
	\label{tab:temperature-dependent_material_properties}
\end{table}

\begin{figure}[H] 
	\centering
	\includegraphics[width=0.75\linewidth]{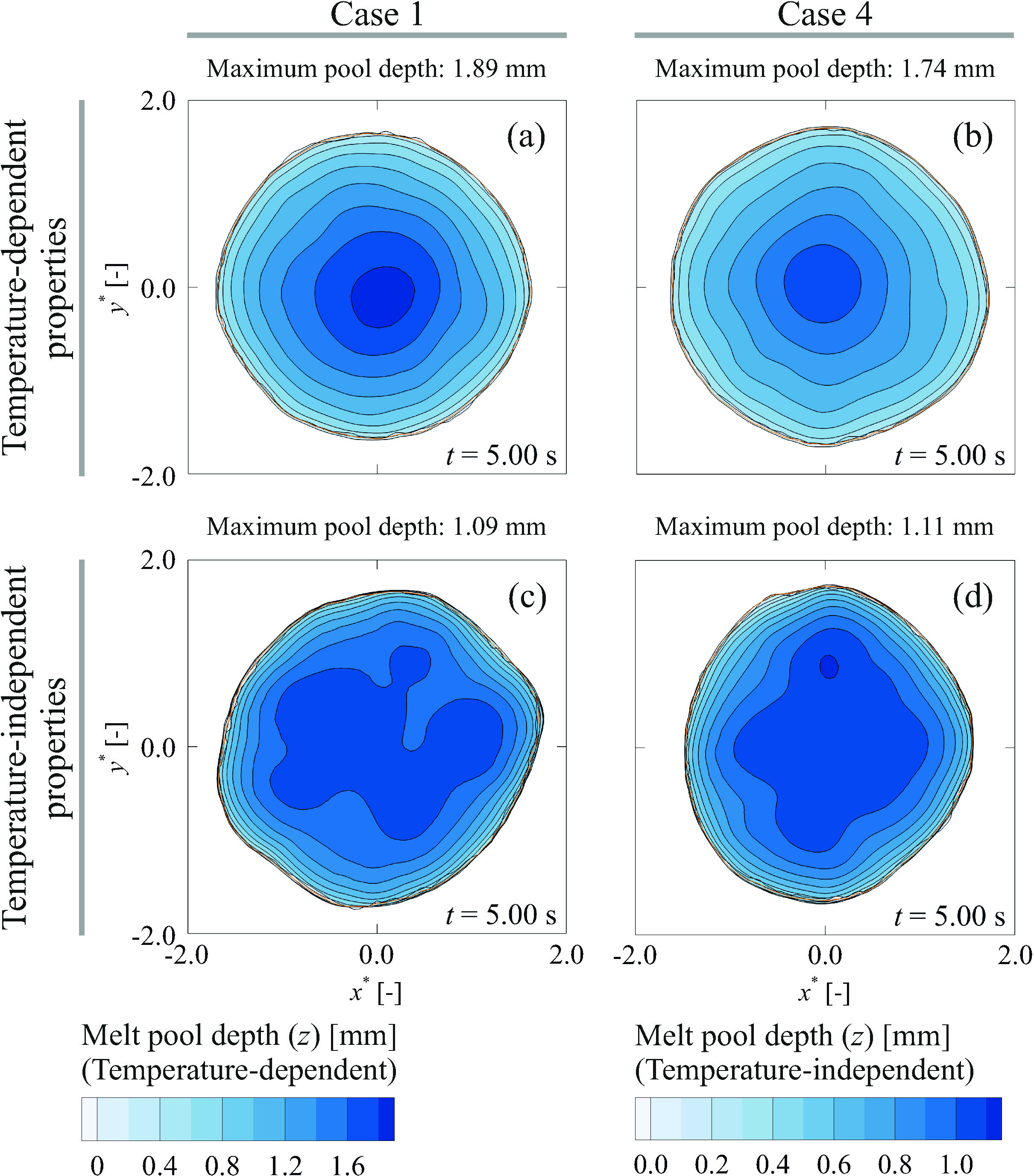}
	\caption{Contours of melt-pool depth at $\SI{5}{\second}$ for Case~1 and 4 predicted using temperature-dependent and temperature-independent material properties. Coordinates are non-dimensionalised using the~laser-beam radius $r_\mathrm{b}$ as the~characteristic length scale. The~orange circular line shows the~melt pool boundary at its top surface.}
	\label{fig:pool_depth_s705}
\end{figure}

Changes in the~material properties with temperature affect thermal and flow fields in the~melt pool, resulting in changes in the~predicted melt-pool shape, as shown in \cref{fig:pool_depth_s705}. When temperature-dependent properties are employed, the~melt-pool depth is significantly (about $36$--$74\%$) larger than that predicted using temperature-independent properties.

\begin{figure}[H] 
	\centering
	\includegraphics[width=1.0\linewidth]{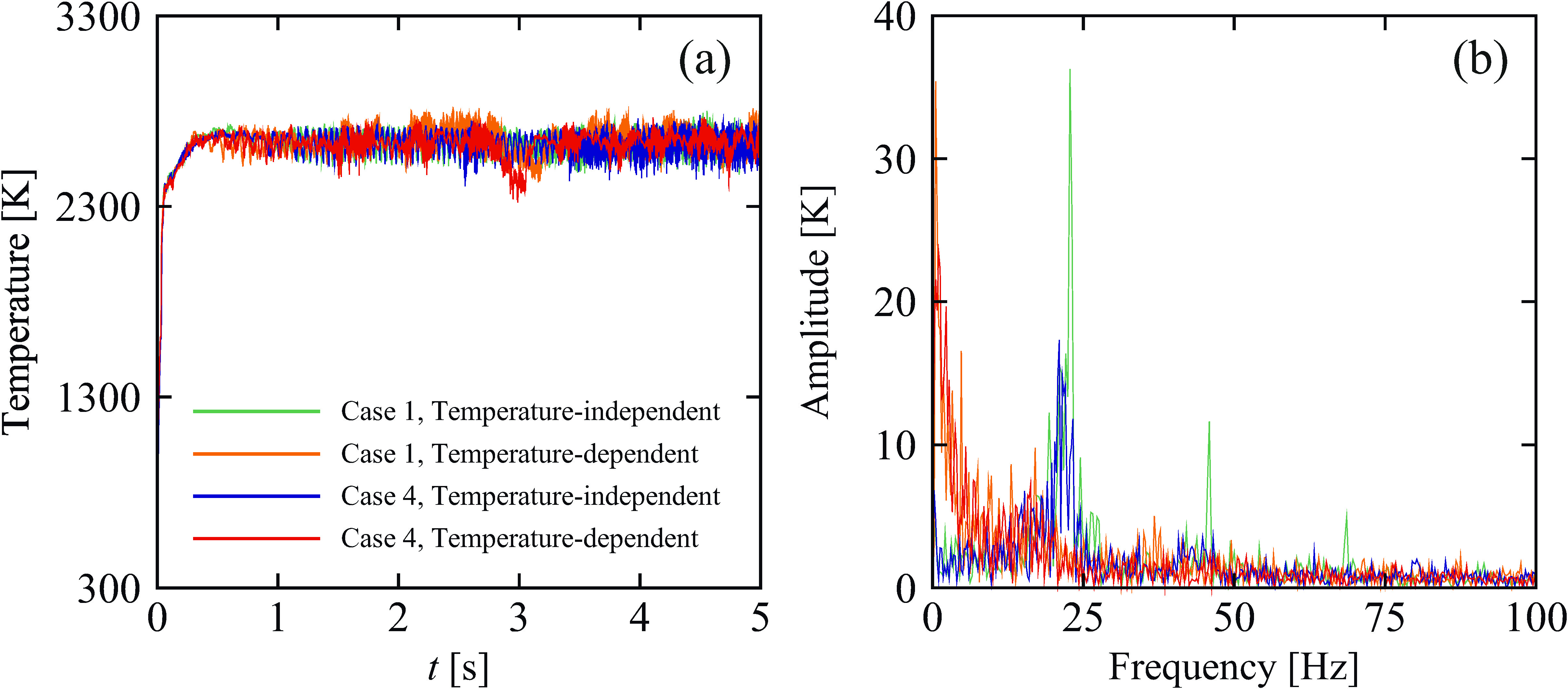}
	\caption{The influence of employing temperature-dependent material properties on temperature signals received from the~melt pool and the~corresponding frequency spectra. (a) Temperature signals recorded from the~monitoring point $p_4$ located at $\left(x^*, y^*\right) = \left(0, 0\right)$ on the~melt-pool surface and (b)~the~corresponding frequency spectra. Temperature signals in the~period of $1$ to $\SI{5}{\second}$ are employed for FFT analysis.}
	\label{fig:temperature_frequency_s705}
\end{figure}

Temperature signals recorded from the~monitoring point $p_4$ and the~corresponding frequency spectra are shown in \cref{fig:temperature_frequency_s705}. Temperature signals received from the~monitoring point $p_4$ are almost in the~same range and vary between $\SI{2320}{\kelvin}$ and $\SI{2820}{\kelvin}$. Employing temperature-dependent properties affects the~frequency spectra of fluctuations, however the~amplitude of fluctuations remains almost unaffected. The~results presented in \cref{fig:temperature_frequency_s705} indicate that adjusting the~heat source dynamically during simulations can result in a~decrease in the~amplitude of temperature fluctuations, however its effect on the~frequency spectrum is insignificant.

\subsection{The effects of the~enhancement factor}
\label{sec:the_effects_of_enhancement_factor}

\begin{figure}[H] 
	\centering
	\includegraphics[width=1.00\linewidth]{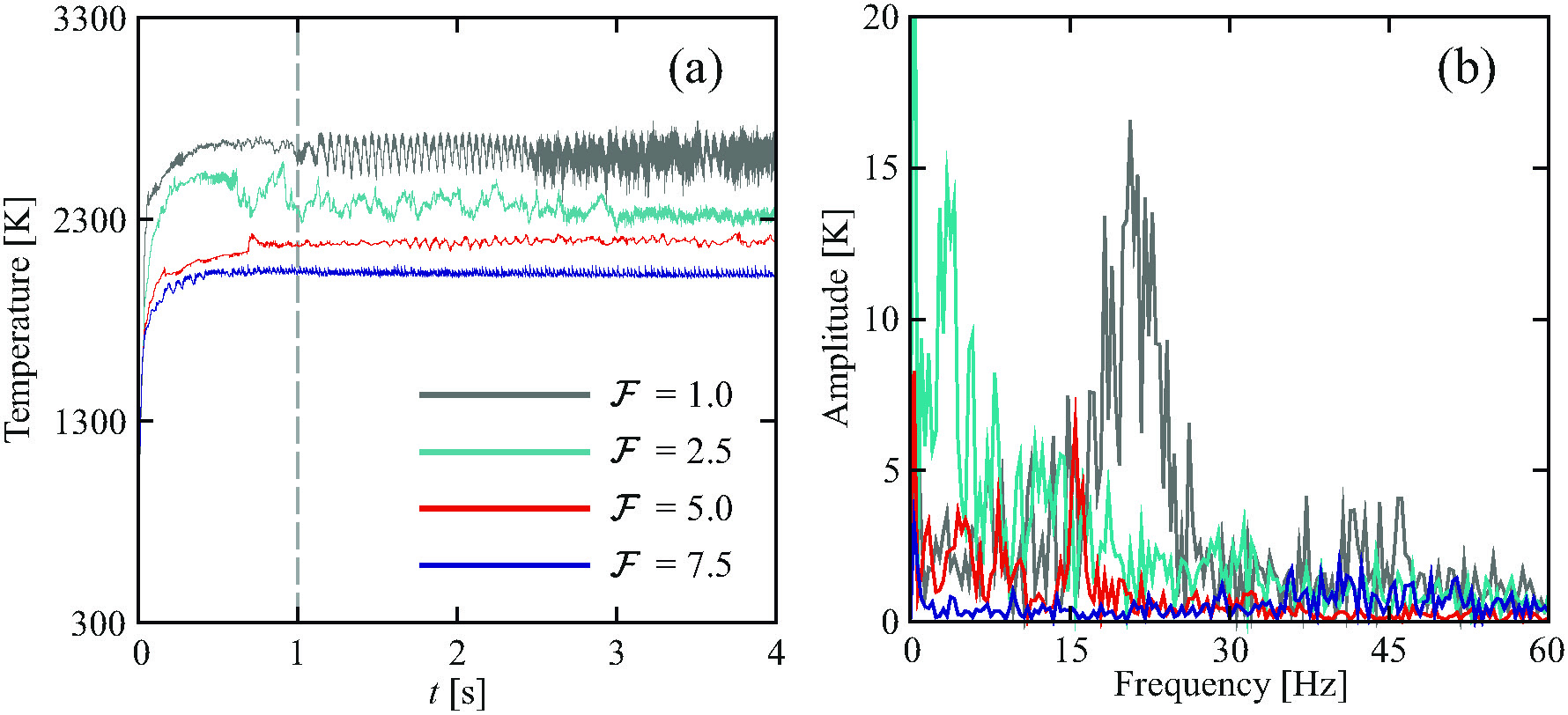}
	\caption{(a)~Temperature signals recorded from the~monitoring point $p_4\left(x^*, y^*\right) = \left(0,0\right)$ on the~melt-pool surface and (b)~the~corresponding frequency spectra for different values of the~enhancement factor $\mathcal{F}$. Temperature signals in the~period of $1$ to $\SI{4}{\second}$ are employed for FFT analysis.}
	\label{fig:enhancement_factor}
\end{figure}

In \cref{sec:validation} it was mentioned that, in many simulation studies in literature, the~thermal conductivity and viscosity of the~liquid metal were artificially increased by a~so-called enhancement factor $\mathcal{F}$, in order to obtain better agreement between experimental and simulated post-solidification melt-pool shapes. Numerical studies carried out by~De \textit{et~al.}~\cite{De_2006,De_2005,De_2004} showed that the~values reported for the~enhancement factor in the~literature depend greatly on operating conditions and ranges from $2$ to $100$ (see for instance,~\cite{Zhang_2003,Choo_1994,Mundra_1993,Choo_1992}), however values between $2$ and $10$ are most often employed. The~effects of the~enhancement factor $\mathcal{F}$ on thermal and fluid flow fields in molten metal melt pools as well as the~melt-pool shape can be found in~\cite{Saldi_2013,Ehlen_2003,Mundra_1992}, thus are not repeated here. Focusing on Case~4 in which free surface deformations are accounted for, the~influence of $\mathcal{F}$ on the~oscillatory flow behaviour is investigated. Temperatures recorded from the~monitoring point $p_4$ and the~temperature fluctuation spectra are shown in \cref{fig:enhancement_factor}. Temperatures and the~amplitudes of fluctuations reduce with increasing $\mathcal{F}$. The~contribution of diffusion in total energy transport increases with $\mathcal{F}$, which results in a~reduction of temperature gradients and therefore thermocapillary stresses generated over the~melt-pool surface. The~reduced thermocapillary stresses in addition to the~enhanced viscosity of the~molten metal lead to a~reduction of fluid velocities that decrease convection in the~melt pool further, which significantly affects the~fluid flow structure. Increasing $\mathcal{F}$ to $2.5$ results in relatively deeper melt pool with a~corresponding reduction of the~fundamental frequency of fluctuations. Using higher values of $\mathcal{F}$, the~melt-pool shape approaches a~spherical-cap shape and the~frequency spectrum becomes rather uniform with small amplitude fluctuations.

\section{Conclusions}
\label{sec:conclusions}

Molten metal melt pool behaviour during a~laser spot melting process was studied to investigate the~influence of dynamically adjusted energy flux distribution on thermal and fluid flow fields using both deformable and non-deformable gas-metal interfaces.

For the~material and laser power studied in the~present work, self-excited flow instabilities arise rapidly and fluid flow inside the~melt pool is inherently three-dimensional and unstable. Flow instabilities in the~melt pool have a~significant influence on solidification and melting by altering the~thermal and fluid flow fields. Free surface deformations, even small compared to the~melt pool size, can significantly influence the~fluid flow pattern in the~melt pool. When in the~numerical simulations the~gas-metal interface is assumed to remain flat and non-deformable, lower temperatures with smaller fluctuations were found in comparison to those of the~cases with a~deformable interface, which results in a~different melt-pool shape. Taking the~surface deformations into account leads to erratic flow patterns with relatively large fluctuations, which are caused by the~intensified interactions between vortices generated in the~melt pool resulting from the~augmented thermocapillary stresses. When surface deformations are taken into account, various tested methods for adjusting the~absorbed energy flux resulted in smaller melt-pool sizes compared to those without an~adjustment. However, the~melt pool behaves quite similarly for various adjustment methods studied in the~present work. This should be noted that the~utilisation of temperature-dependent properties can enhance the~accuracy of numerical predictions in simulations of molten metal flow in melting pools; however, the~results presented in the~present work show the~importance of employing a~physically-realistic heat-source model that is also applicable if temperature-dependent properties were employed.

Although the~enhancement factors are widely used to achieve agreement between numerically predicted melt-pool sizes and solidification rates with experiments, they do not represent the~physics of complex transport phenomena governing laser spot melting. The~use of an~enhancement factor can significantly affect the~numerical predictions of melt pool oscillatory behaviour.

\section*{Author Contributions}
\label{sec:author_contributions}

Conceptualisation, A.E., C.R.K. and I.M.R.; methodology, A.E.; software, A.E.; validation, A.E.; formal analysis, A.E.; investigation, A.E.; resources, A.E., C.R.K, and~I.M.R.; data curation, A.E.; writing---original draft preparation, A.E.; writing---review and editing, A.E., C.R.K.,  and~I.M.R; visualisation, A.E.; supervision, C.R.K. and  I.M.R.; project administration, A.E. and  I.M.R.;  and funding acquisition, I.M.R. 

\section*{Acknowledgement}
\label{sec:acknowledgement}

This research was carried out under project number F31.7.13504 in the~framework of the~Partnership Program of the~Materials innovation institute M2i (www.m2i.nl) and the~Foundation for Fundamental Research on Matter (FOM) (www.fom.nl), which is part of the~Netherlands Organisation for Scientific Research (www.nwo.nl). The~authors would like to thank the~industrial partner in this project “Allseas Engineering B.V.” for the~financial support.

\section*{Conflict of interest}
\label{sec:conflict_of_interest}

The authors declare no conflict of interest.

\appendix

\section*{Appendix}
\section{Grid independence test}
\label{app:grid_study}

\begin{figure}[H]
	\centering
	\includegraphics[width=0.60\linewidth]{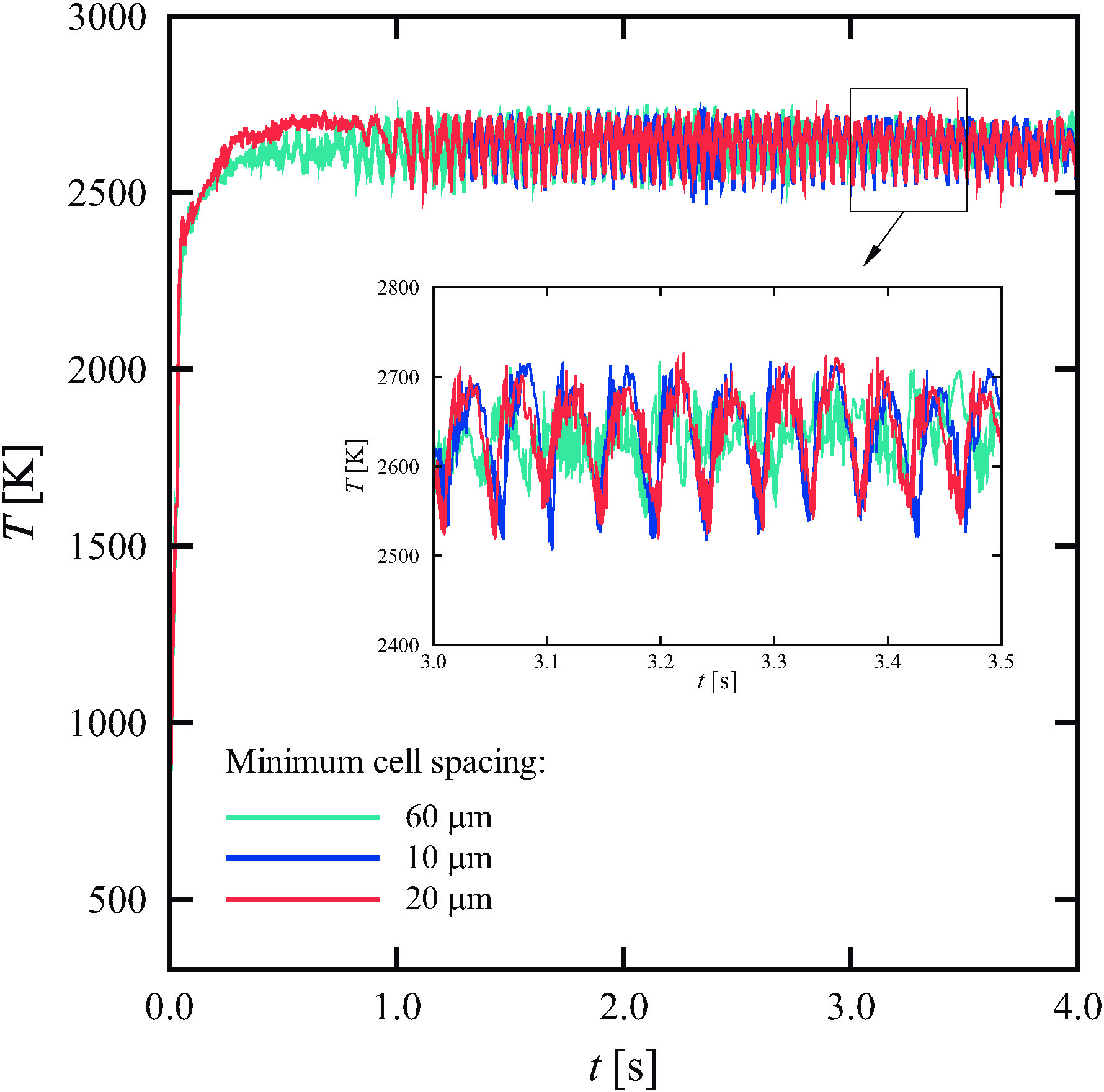}
	\caption{Temperature variations predicted at the~melt pool centre using different grid sizes. The~grid with minimum cell spacing of $\SI{20}{\micro\meter}$ was employed for the~calculations reported in the~paper.}
	\label{fig: grid_study}
\end{figure}

\begin{figure}[H]
	\centering
	\includegraphics[width=0.50\linewidth]{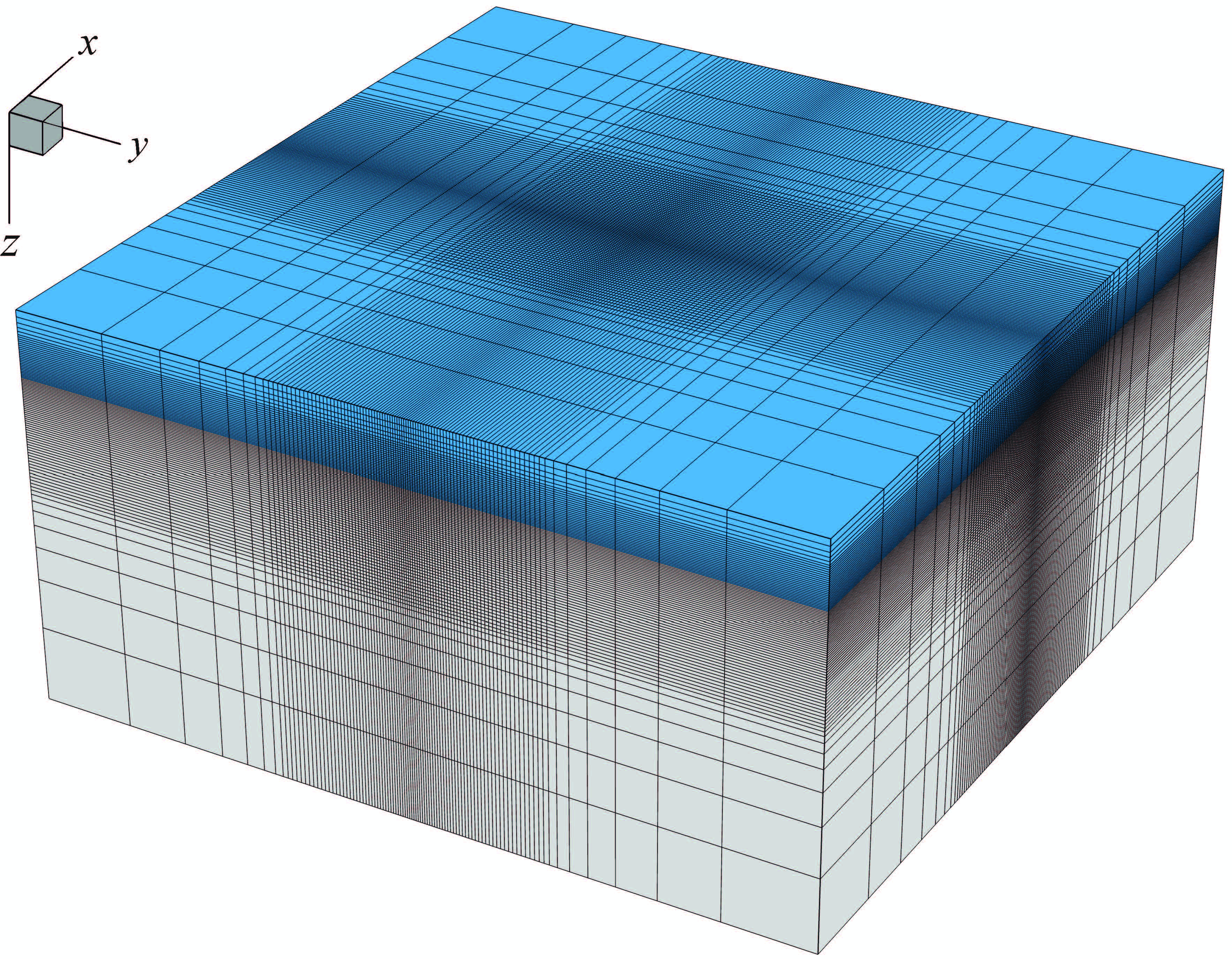}
	\caption{The computational grid employed for the~present calculations reported in the~paper. Regions highlighted in blue show the~gas layer above the~base material.}
	\label{fig: grid}
\end{figure}

Case 1 was considered for grid independence test. Three different grids with minimum cell spacings of $60$, $20$ and $\SI{10}{\micro\meter}$ were studied. The~influence of computational cell size on predicted melt pool depth and width are reported in \cref{tab: grid_study}, which indicate the~predictions are reasonably independent of the~grid size. The~variations of temperature at the~centre of the~melt pool surface (\textit{i.e.} $T(0, 0, z_\mathrm{surface})$) were also investigated for different grid sizes and the~results are shown in \cref{fig: grid_study}. The~grid with minimum cell size of $\SI{20}{\micro\meter}$ was employed for the~present calculations reported in the~paper, and is shown in \cref{fig: grid}.

\bgroup
\def\arraystretch{1.15}	
\begin{table}[h] 
	\centering
	\caption{The influence of computational cell size on predicted melt pool size.}
	\begin{tabular}{llll}
		\hline
		Minimum cell spacing    & Total number of cells & Melt pool width           & Melt pool depth           \\ \hline
		$\SI{60}{\micro\meter}$ & $\SI{5.5e5}{}$          & $\SI{4.89}{\milli\meter}$ & $\SI{1.09}{\milli\meter}$ \\
		$\SI{20}{\micro\meter}$ & $\SI{8.6e5}{}$          & $\SI{4.88}{\milli\meter}$ & $\SI{0.96}{\milli\meter}$ \\
		$\SI{10}{\micro\meter}$ & $\SI{2.1e6}{}$          & $\SI{4.88}{\milli\meter}$ & $\SI{0.95}{\milli\meter}$ \\ \hline
	\end{tabular} 
	\label{tab: grid_study}
\end{table}
\egroup

\small
\bibliographystyle{elsarticle-num}
\bibliography{ref}

\end{document}